\documentclass[times,review,preprint,authoryear]{elsarticle}

\usepackage{jasr}
\usepackage{framed,multirow}

\usepackage{amssymb}
\usepackage{latexsym}

\usepackage{url}
\usepackage{xcolor}
\definecolor{newcolor}{rgb}{.8,.349,.1}

\usepackage[citebordercolor=white]{hyperref}

\journal{Advances in Space Research}

\begin{document}

\verso{Given-name Surname \textit{etal}}

\begin{frontmatter}

\title{Robustness analysis and station-keeping control of an interferometer formation flying mission in low Earth orbit}%

\author[1]{Cristina \snm{Erbeia}}
\ead{cristina.erbeia@mail.polimi.it}
\author[1]{Francesca \snm{Scala}\corref{cor1}}
\cortext[cor1]{Corresponding author}
\ead{francesca1.scala@polimi.it (Currently at German Aerospace Center (DLR), Oberpfaffenhofen, Germany)}
\author[1]{Camilla \snm{Colombo}}
\ead{camilla.colombo@polimi.it}

\address[1]{Aerospace Science and Technology Department, Via La Masa 34, Politecnico di Milano, Milano, 20156, Italy}


\begin{abstract}
The impact of formation flying on interferometry is growing over the years for the potential performances it could offer. However, it is still an open field, and many studies are still required.
This article presents the basic principles behind interferometry focusing first on a single array and secondly on a formation of satellites. A sensitivity analysis is carried out to evaluate how the performance of the interferometry is affected by an error in the relative position in the formation geometry. This is estimated by computing the loss of the performances in terms of percentage deviation due to a non-nominal relative trajectory, including two-dimensional errors and defining a payload index. The main goal of this study is to estimate whether some errors in the relative state are more impacting than others. The final objective is to compute the link between a certain error and a specific loss of performance, to eventually foresee where the error is found. Furthermore, a dynamical model is developed to describe the relative motion in the Low Earth Orbit environment, considering both the unperturbed and the $J_{2}$ and drag scenarios. A Proportional, Integral and Derivative controller is implemented for the position control of a multiple satellite formation flying, considering a low thrust control profile. The FFLAS study is taken as a reference configuration, analysing both nominal and non-nominal configurations. 
This study could serve as a starting point for the development of a combined tool to assess the performances of both the interferometry and the control on the relative state for future remote sensing studies involving relative motion.
\end{abstract}

\begin{keyword}
\KWD FFLAS \sep formation flying \sep interferometry \sep payload modelling \sep formation control.
\end{keyword}

\end{frontmatter}


\section{Introduction}
In recent years, interferometry formation flying is becoming more and more relevant for Earth monitoring, as it could improve the performance of a radiometer keeping a reasonable satellite size. Passive radiometers are the most used technique \citep{Erbeia2022, Kerr2010, AGU} to detect both the soil moisture and ocean salinity with a single instrument, particularly the L-band radiometer. They have, however, an intrinsic constraint: the spatial resolution is directly proportional to the antenna diameter and inversely to the wavelength. Consequently, large antenna apertures are required to improve spatial resolution. The detection of soil moisture and ocean salinity puts some constraints on the accuracy, spatial resolution and temporal sampling. To fulfil the requirements without deploying a large array, a synthesised antenna could be implemented \citep{Neira2022, Zurita2013}. This latter option is known as a microwave interferometric radiometer, and it is selected, for instance, in the SMOS mission carrying the MIRAS payload \citep{Kerr2010}. However, the SMOS mission has left open many challenges, such as the spatial resolution refinement, the temporal resolution improvement, the implementation of a device with a cost-effective design or the robustness to radio frequency interferences \citep{AGU}.
ESA and CNES are facing these aspects. In particular, the ESA's SMOSops deals with all the mentioned aspects but the improved spatial resolution, which is still comparable to the SMOS (i.e. around 40 $km$ resolution). The CNES SMOS-NEXT instead investigates spatial resolution and sensitivity, by combining spatial and temporal 2D interferometry \citep{AGU}. To improve even further the resolution of a radiometer, a fleet of satellites could be implemented, as setting a formation of satellites allows enlarging the satellites' coverage without augmenting too much the array size of a single satellite \citep{Scala2020, Neira2022}. Therefore, formation flying is promising in Earth observation for the performance improvements it could offer.\\

In this work, both the payload performances and the control for the formation maintenance are investigated.  As a result, they not only interface to provide input data, but also merge for a more critical and optimized overall analysis. Improving the payload optimization means indeed maximising the performances, a critical aspect of the space economy and exploration \citep{Zurita2013, AGU}. Some novelties are introduced for the payload modelling, such as for the array factor. In the array factor computation (Eq. \ref{eq:AF}), each \textcolor{black}{\textit{(u,v)}} couple \textcolor{black}{(i.e. each cartesian point \textcolor{black}{\textit{(x,y)}} normalized by the wavelength $\lambda$)} must be counted once, without considering the redundancy points. An inaccurate \textcolor{black}{\textit{(u,v)}} points count leads to a bad interpretation of the actual sidelobes.
This aspect is faced through two methodologies. In the first case, the unique \textcolor{black}{\textit{uv}} choice is \textcolor{black}{computed via an already available algorithm in the literature. On the contrary, in the second case, a new methodology has been developed, called parallelogram method. In this approach, the \textit{uv}} count is made through some simple geometrical assumptions, taking inspiration from \cite{Grzesik2006}. This latter can be considered a valid comparison to more classical approaches. The payload performances are evaluated by understanding the link between the relative position error and the array factor. The analysis considers errors varying in magnitude, quantity and direction, and it is carried out both from a quantitative and qualitative point of view. The deviations from the nominal performance are quantified, and a performance payload index is defined. Finally, the analysis is extended by considering a close loop control for the assessment of the formation maintenance and the evaluation of the control error on the relative position. The control part is analysed not only from the position point of view but also from the modelling one. 
The collected results allow both to expand the modelling evaluations and link the modelling and the control part. This provides the foundations of a machine learning-like approach, as an alternative to the classical physics-based models, for increasing the satellites' position accuracy.

\section{Interferometric radiometry for multiple satellites}
\label{sec:interferometry}
This section presents the fundamental theory behind interferometry applied in \cite{Erbeia2022}, and derived by the work in \cite{Thompson2017} and \cite{Ahmed2019}.
It describes the main results from the Super-MIRAS case study \citep{Erbeia2022, Zurita2013}, and it finally presents the formation flying geometry analysed in this article: Formation Flying L-band Aperture Synthesis (FFLAS) study from the work in \cite{Scala2020}. 

The main goal of this work is to check how the payload performances are affected by a position error in the formation geometry. Consequently, a  special focus is made on the array factor, i.e. a direct indicator of how the array is performing in terms of image quality \citep{Wu2017}.

\subsection{Spatial frequency coverage} \label{subs:SFC}
Starting from the findings in \cite{Ahmed2019}, an array of antenna elements composing the interferometer is necessary to remove the ambiguities in the observation of multiple sources. This distribution of antenna elements is called Array Configuration (AC), and its output is called Visibility function \textcolor{black}{\textit{V}}. The latter is a complex quantity that can be defined from the Fourier transform of the source intensity \textcolor{black}{\textit{B}} (or brightness map) \citep{Thompson2017, Ahmed2019}. This is defined as the amount of energy emitted by the source observed by the interferometer. As described in \cite{Ahmed2019}, the set of visibility functions from the array configurations defines the visibility sampling. 
Similarly to the analyses in \cite{Ahmed2019}, we introduce two coordinate systems: the cartesian coordinate system \textcolor{black}{\textit{(x,y,z)}} to define the geometry and the configuration of the arrays, and the coordinate system \textcolor{black}{\textit{(u,v,w)}} to characterise the visibility function \textcolor{black}{\textit{V}}. The \textcolor{black}{\textit{(u,v,w)}} reference frame is defined from the cartesian one, by normalising the \textcolor{black}{\textit{(x,y,z)}} directions over the antenna wavelength lambda, \textcolor{black}{where the \textit{uvw} points represent the visibility points related to the visibility function \textit{V}}.

Starting from the Visibility \textit{\textbf{V}} definition and applying some assumptions as explained in \cite{Ahmed2019}, the so-called Van Cittern-Zernike formulation is obtained as in Eq. \ref{eq:VCZ} \citep{Ahmed2019, Thompson2017}. This formulation describes that the visibility function \textcolor{black}{\textit{V}} referred to the observed brightness map \textcolor{black}{\textit{B}}, \textcolor{black}{\textit{$V_{B}$(u,v)}}, is equivalent to the two-dimensional spatial Fourier transform of $\Tilde{B}(\xi,\eta)$ = $B(\xi,\eta)$ / $\sqrt{1-\xi^{2}-\eta^{2}}$, the so-called modified brightness map:

\begin{equation} \label{eq:VCZ}
    V_{B} (u,v) = \int\int_{\xi^2+\eta^2\leq 1} \frac{B(\xi,\eta)}{\sqrt{(1-\xi^{2}-\eta^{2})}} e^{-2i\pi(u\xi +v\eta)} d\xi d\eta,
\end{equation}
where $\xi$ and $\eta$ are the direction cosines which represent the direction of the source \textcolor{black}{(being the frame of the direction source $\hat{s}$ = $\begin{bmatrix}{ \xi; \eta; \sqrt{1- \xi^{2} - \eta^{2}}} \end{bmatrix}$, \cite{Ahmed2019}) }, and i represents the complex quantity $\sqrt{-1}$.

As in a Fourier transform the input is in the spatial domain while the output is in the frequency domain, the \textcolor{black}{\textit{(u,v)}} points represent the spatial frequency components in $\xi$ and $\eta$. 
For this reason, the visibility points \textcolor{black}{\textit{uv}} are also referred to as Spatial Frequency Coverage (SFC). By increasing the \textcolor{black}{\textit{(u,v)}} coverage and decreasing the spatial period, the spatial resolution improves, as the \textcolor{black}{\textit{(u,v)}} points can detect finer details in the initial brightness map \citep{Ahmed2019}. A more optimized SFC means therefore better performances. For this reason, different geometries are studied \citep{Zurita2013, Ahmed2019, Erbeia2022}, and are further analysed in sections \ref{subs:results1} and \ref{subs:FFLAS}. The SFC is also the starting point of the sensitivity analyses presented in section \ref{sec:poserror}. 

\subsection{Interferometer's array factor} \label{subs:AF}
First, we determine the Array Factor \textcolor{black}{\textit{AF}} in the ($\xi$,$\eta$) plane, and then we express the sidelobe levels through the quantity of interest in \textcolor{black}{$dB$} (Eq. \ref{eq:SLL}), as suggested by \cite{Woodhouse2006}. Considering an antenna, the sidelobes represent smaller peaks with lower gains which are translated from the central, maximum peak.
The Array Factor \textcolor{black}{\textit{AF}} can be computed inverting the visibility equation via inverse Fourier transform \citep{Ahmed2019}. Being the \textcolor{black}{\textit{(u,v)}} points not continuous but discretised samples, the \textcolor{black}{\textit{AF}} definition is expressed as a discretised summation, as in Eq. \ref{eq:AF} for the ideal case, i.e. as expressed by the assumptions introduced in \cite{Ahmed2019} \citep{Wu2017} . 


\begin{equation} \label{eq:AF}
  AF(\xi,\eta,\xi_{0}, \eta_{0}) = \sum_{m=1}^{N} W_{m}e^{2\pi i(u_{m}(\xi-\xi_{0})+v_{m}(\eta-\eta_{0}))} \Delta s_{m}, 
\end{equation}

\begin{equation} \label{eq:SLL}
    \textrm{Quantity of Interest in [dB]} = 10log_{10}\frac{\textrm{Quantity of Interest}}{\textrm{Reference Quantity}},
\end{equation}
with \textcolor{black}{\textit{N}} visibility samples and \textcolor{black}{\textit{W}} related weighting functions. $\Delta s_{m} $ is the sampling points occupation area, and it depends on the considered array. \textcolor{black}{The sampling point occupation area represents the area that the \textit{uv} points cover in the sampling grid where the image reconstruction process is performed. For instance, for a hexagonal array, this value is equal to $\frac{\sqrt{3}}{2}d^{2}$ being \textit{d} the spacing factor (see Table \ref{tab:FFLAS_characteristics} as an example)}. In this case, the Quantity of Interest is the array factor \textcolor{black}{\textit{AF}} for each point of the ($\xi$,$\eta$) grid, while the Reference Quantity is the maximum array factor value corresponding to the source location point ($\xi_{0}$,$\eta_{0}$). The Quantity of Interest in [dB] is, therefore, the sidelobe level (\textcolor{black}{\textit{SLL}}).

\begin{figure}[H]
\centering
\subfloat[]
{
     \includegraphics[scale = 0.3]{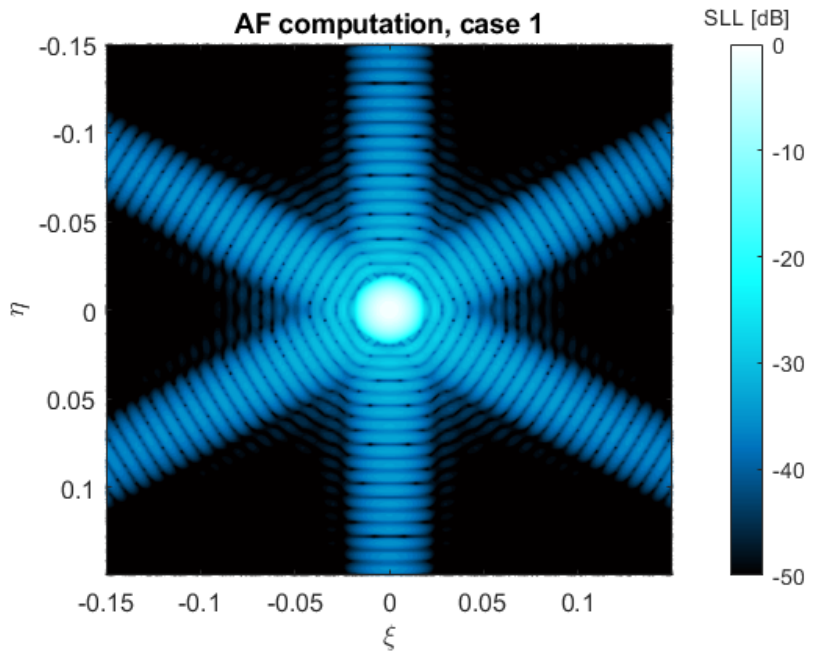}
     \label{fig:case1hexagon}
     }
     \subfloat[]
{
    \includegraphics[scale = 0.3]{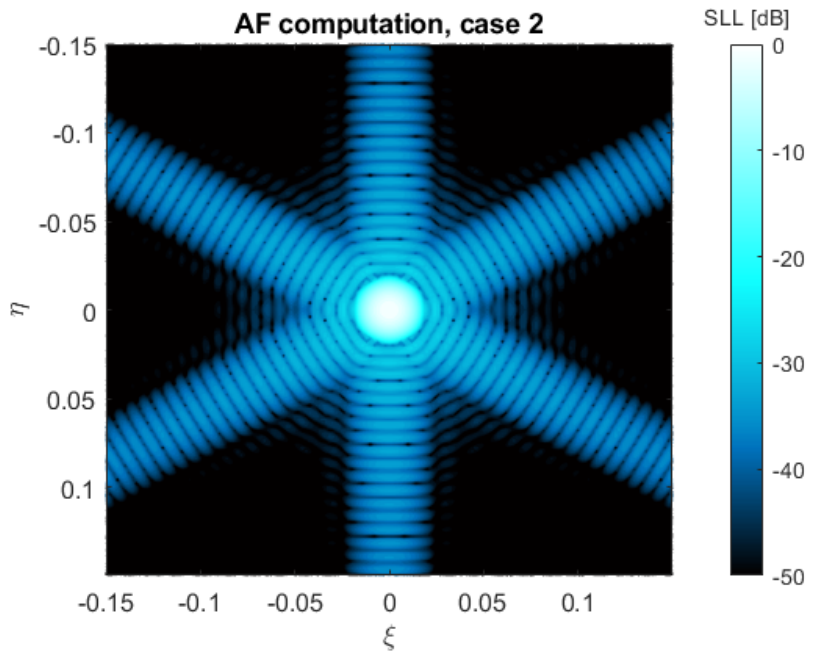}
     \label{fig:case2hexagon}
     }
   
 \subfloat[]{    
      \includegraphics[scale = 0.3]{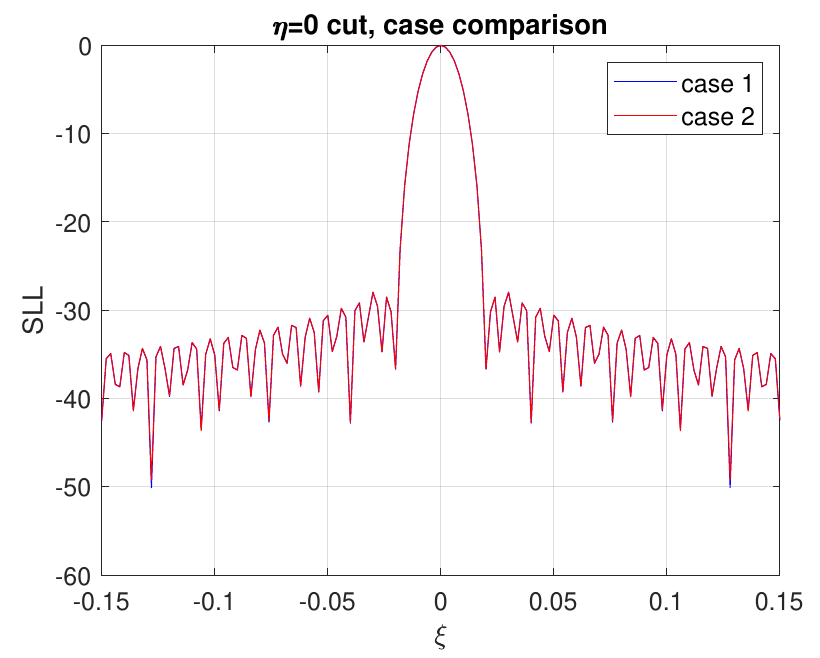}
     \label{fig:xiplane_comparison}
}
\subfloat[]
{     
 \includegraphics[scale = 0.3]{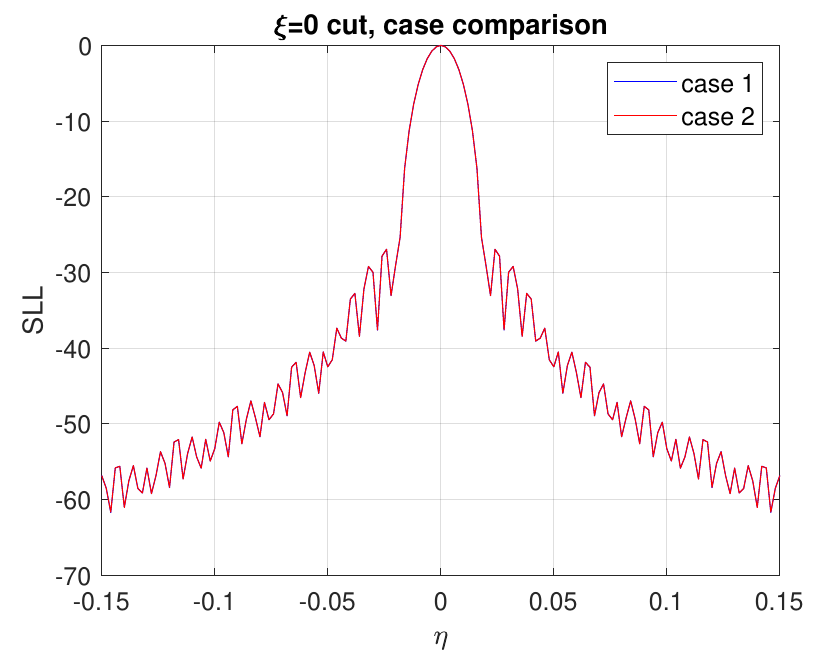}
     \label{fig:etaplane_comparison}
     }
      \caption{Comparison of array factors and sidelobe levels (\ref{fig:xiplane_comparison} and \ref{fig:etaplane_comparison}) for the Matlab method (\ref{fig:case1hexagon}) and the parallelogram method (\ref{fig:case2hexagon}) \citep{Erbeia2022}.}
        \label{fig:SLL_comparison}
\end{figure}

The array factor could be computed through two different methods, called the Matlab method and the parallelogram method, as described in a previous work \citep{Erbeia2022}.  The main idea is to optimise the \textcolor{black}{\textit{AF}} and redundancy point computation by investigating different solutions, and, at the same time, to validate the obtained results through a case-to-case comparison. In the former, the \textcolor{black}{\textit{(u,v)}} points are counted once from the original spatial frequency coverage, returning each unique \textcolor{black}{\textit{uv}} point but a tolerance (therefore, deleting the redundancy points). In the parallelogram method the lattice is reconstructed, subdividing the spatial frequency coverages in parallelograms and deleting the redundancy points in between \citep{Grzesik2006}, following the Algorithm described in \cite{Erbeia2022} (Algorithm \ref{alg:var}). The Matlab method utilises some automated commands. The \textcolor{black}{\textit{uv}} set is therefore selected by the computer, and the process is not directly controlled. The parallelogram method allows the user to be more aware of how to get an array factor, and control the selection of the \textcolor{black}{\textit{uv}} redundancy points.
The two methods give similar results: the mean deviation for the sidelobe results between the two cases is 0.11\% for an hexagonal array, as in  Fig. \ref{fig:SLL_comparison} \citep{Erbeia2022}. While the first method is quicker to apply, the parallelogram one could be used to verify and validate the process from the Matlab method. \textcolor{black}{ For instance, it has been verified that the two methods lead to an exact count of the \textit{uv} points for an open geometry, whereas to a slightly higher and inexact for the Matlab method when close geometries are analysed. This difference could be further inspected in future analyses.}

\begin{algorithm}[H]
    \label{alg:example}
    \caption{Parallelogram method}
    \label{alg:var}
    \label{protocol1}
    \begin{algorithmic}[1]
    \STATE Define AC and SFC. Through SFC, define the reference points for drawing the parallelograms, i.e. the vertices of the fundamental hexagon and the point (0,0). From the reference points, define one reference vector for each parallelogram, i.e. one side for each parallelogram. They should all have the same length.
    \FOR{$n = 1: length(reference$ $vector)$}
    \STATE{draw parallel vectors to the reference ones until all the parallelograms are drawn}
    \ENDFOR
    \STATE Delete redundancy points in between the parallelograms
    \end{algorithmic}
\end{algorithm}

\subsection{Redundancy factor} \label{subs:redundancy}
Considering the array factor in Eq. \ref{eq:AF}, each couple \textcolor{black}{\textit{(u,v)}} must be counted only one time, independently from the order, removing the redundancy in the computation.

The Spatial Frequency Coverage must in any case be guaranteed: a hole in the \textcolor{black}{\textit{(u,v)}} plane means a deterioration in the image quality \citep{Zurita2013}. The redundancy points are therefore effective in describing how robust is the interferometer against failures, through the so-called Redundancy Factor (\textcolor{black}{\textit{RF}}):

\begin{equation} \label{eq:RF}
    RF= \sqrt{\frac{N'_{v}}{N_{v}}},
\end{equation}

where $N_{v}$ is the number of \textit{(u,v)} points without considering their Redundancy Order (\textcolor{black}{\textit{RO}). The Redundancy Order is the number of times a \textit{uv} couple is obtained from its computation, while the Redundancy Factor compares the \textit{uv} points considering their Redundancy Order $N'_{v}$ (i.e. $N'_{v}$ =  $\sum_{u,v}$ $1$ / $RO$), to $N_{v}$}. The lower the redundancy factor, the higher is the robustness of the system \citep{Zurita2013}.

\subsection{Windowing functions} \label{subs:windowing}
The \textcolor{black}{\textit{W}} quantity in Eq. \ref{eq:AF} is the window function, a mathematical function that values zero outside of some selected intervals.
There are different metrics to select the appropriate window, such as the main lobe width and the sidelobe peak level. The first is referred to spatial resolution, while the second one to radio frequency interferences mitigation. A third metric is the value at 0.5 of the frequency, which is called the maximum scalloping loss of the window. 

The window selection is therefore the result of a performance tradeoff. 
Table \ref{tab:windowing} shows the most common windowing in signal processing, i.e. Rectangular, Triangular, Hamming, Hanning and Blackman, which is the one used for the SMOS mission. The other windows are introduced for comparison and for validating the choice performed for SMOS.

\begin{table}[H]
\centering
\caption{Selected windowing functions \citep{Ahmed2019},}
\begin{tabular}{|p{1.8 cm} | p{6 cm} |} 
\hline
 \rowcolor[HTML]{C0C0C0}
\textbf{Windowing} & \textbf{Definition} \\ \hline \hline
 \textbf{Rectangular} & W=1 \\ \hline
  \textbf{Triangular} & W= 1 - $\frac{\Tilde{\rho}}{\rho_{max}}$ \\ \hline
  \textbf{Hanning} &  W= $0.5+0.5\cos\bigg( {\pi\bigg(\frac{\Tilde{\rho}}{\rho_{max}}\bigg)}\bigg)$ \\ \hline
  \textbf{Hamming} &  W= $0.54+0.46\cos\bigg( {\pi\bigg(\frac{\Tilde{\rho}}{\rho_{max}}\bigg)}\bigg)$\\ \hline
 \textbf{Blackman} &  W= $0.42+0.5\cos\bigg( {\pi\bigg(\frac{\Tilde{\rho}}{\rho_{max}}\bigg)}\bigg)+0.08\cos\bigg({2\pi\bigg(\frac{\Tilde{\rho}}{\rho_{max}}\bigg)}\bigg)$   \\ \hline

\end{tabular}
\label{tab:windowing}
\end{table}

where $\Tilde{\rho}$ = $\sqrt{u^{2}+v^{2}}$ is the distance radius from the origin for each pair of antennas and $\rho_{max}$ is its maximum value (\cite{Ahmed2019}).

The analysis performed in \cite{Erbeia2022} shows that the Blackman window is the best at reducing the sidelobe level effect. For this reason, this window is implemented in the case studies of this article. 
The downside of the Blackman window is the main lobe broadening, which results in a spatial resolution worsening.  

\subsection{Alias-free field of view} \label{subs:AF-FOV}
The set of visibility samples and the spacing between antennas are related to the instrument's field of view, as described in \cite{Duran2017}. Considering an antenna spacing higher than the Nyquist sampling rate value, aliases appear in the periodic extension. The area not subjected to aliases is defined as the Alias-free field of view, and it is inversely proportional to the antenna spacing. Therefore, reducing the antenna spacing means getting a larger alias-free field of view, so a larger swath, as for the Super-MIRAS case study \citep{Zurita2013}. This latter value is quite important, as all three basic science requirements, i.e. the resolution, revisit time and accuracy, can be expressed in terms of maximum instrument swath. 

\subsection{Performance of the single array configuration} \label{subs:results1}
\begin{figure}[H]
\centering
\subfloat[]
{
     \includegraphics[scale = 0.3]{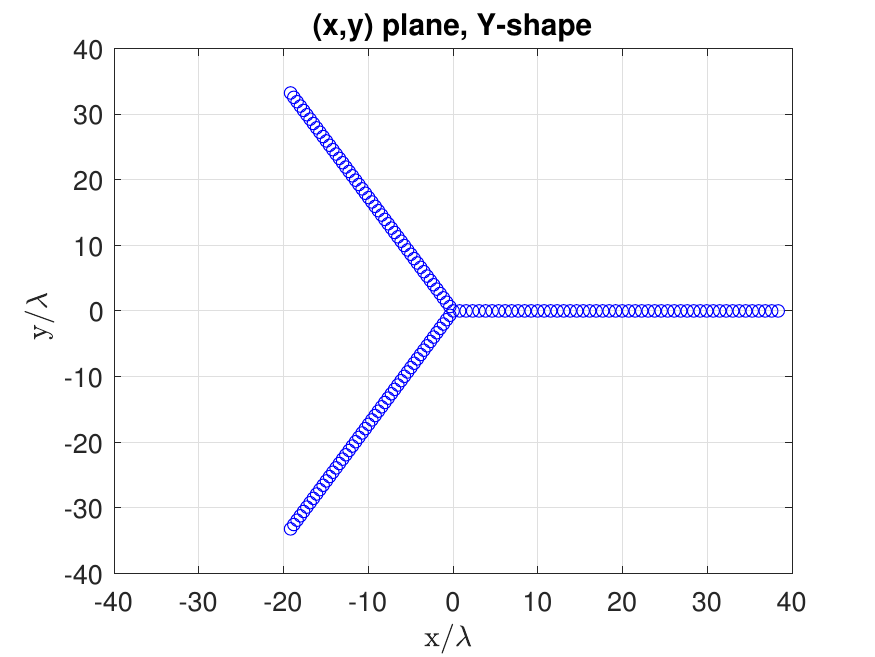}
     }
    \subfloat[]{    
      \includegraphics[scale = 0.3]{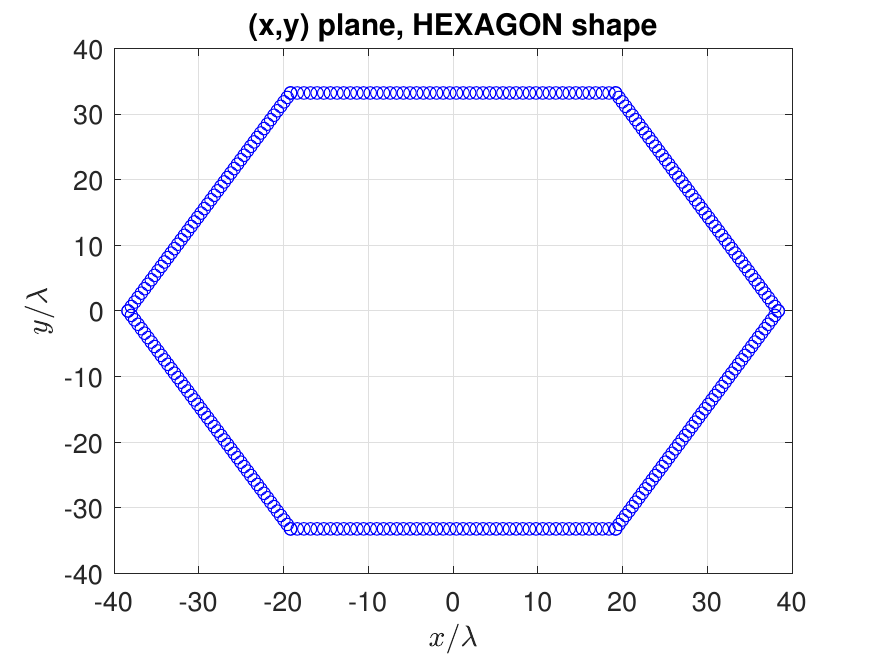}
}
   
     \subfloat[]
{
    \includegraphics[scale = 0.3]{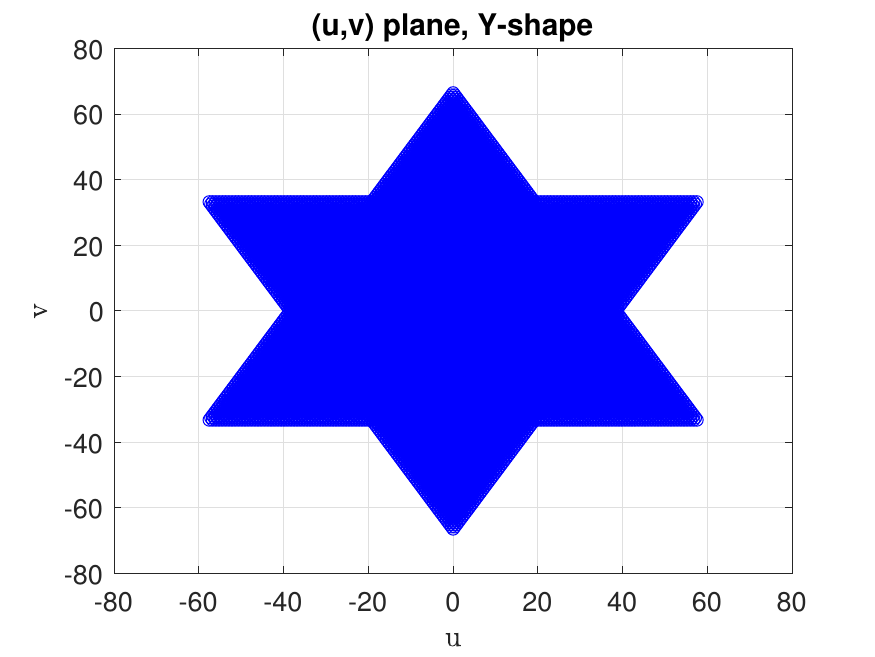}
     }
\subfloat[]
{     
 \includegraphics[scale = 0.3]{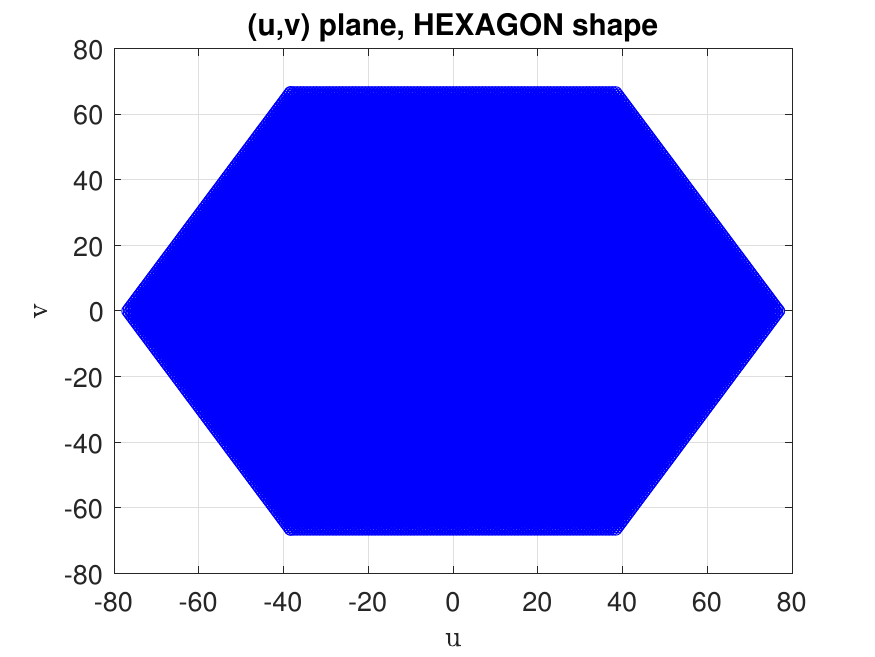}
     }

     \subfloat[]
{
    \includegraphics[scale = 0.32]{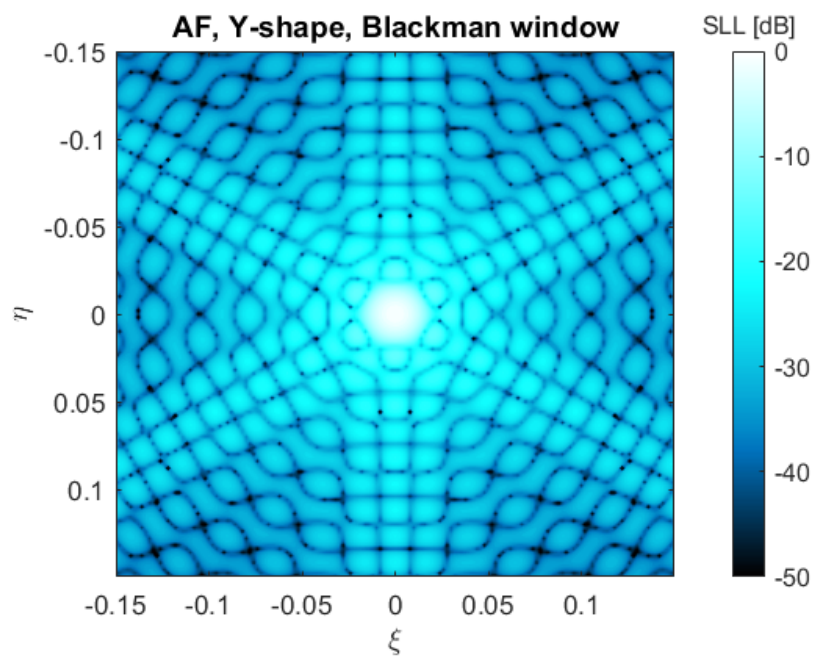}
     }
\subfloat[]
{     
 \includegraphics[scale = 0.32]{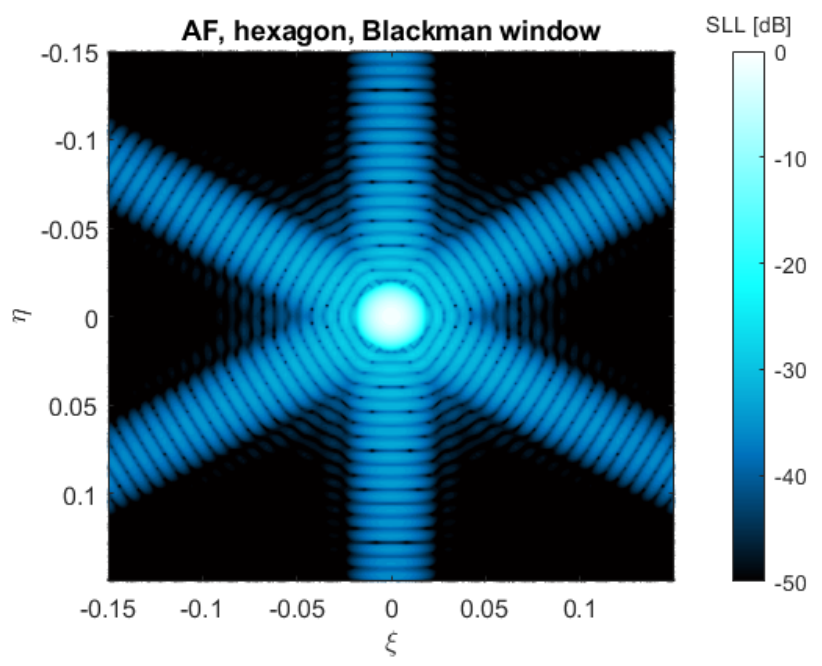}
     }
   \caption{Geometries (up), coverages (middle) and array factors (down) for the y and hexagon array from the Super-MIRAS case-study.}
        \label{fig:all_SuperMIRAS}
    
\end{figure}

All the principles previously listed are first applied to single array configurations \citep{Erbeia2022}, e.g. y, star, double-y, triangular, and hexagon array, taking inspiration from the Super-MIRAS study described in \cite{Zurita2013}. Fig. \ref{fig:all_SuperMIRAS} shows as example a comparison between the y shape and the hexagon array. The following considerations arised:
\begin{itemize}
   \item lowering the antennas spacing allows to enlarge the alias-free field of view;
    \item a hexagonal array has the largest spatial frequency coverage, a double-y one the thinnest;
    \item a star array has the lowest redundancy factor, a y-shape the highest. The hexagonal one is a good compromise;
    \item a close array leads to a more effective sidelobe level reduction than an open array;
    \item the Blackman window well reduces the sidelobe levels, at the cost of the main lobe broadening.
\end{itemize}

For these reasons, a hexagon array is the most suitable one for reducing the sidelobe levels, having a good degree of robustness, and improving the interferometer performances.

\subsection{Performance of multiple arrays configuration} \label{subs:multiplearrays}
A potential way to refine the spatial resolution is to move from a single satellite to a formation. Setting a formation of satellites allows to broaden the spatial frequency coverage without increasing too much the array size of a single satellite \citep{Zurita2013, Neira2022}. In addition to the geometric characteristics, such as the number of antennas or the antenna spacing, also the relative position between the satellites plays an important role.
Satellites in formation can be arranged in multiple configurations. Following the analyses in \cite{Ahmed2019}, the main dispositions and formations could be respectively single-element companion and array duplicate, and planar, planar tilted, and staggered arrays.

Starting from the analysis done in \citep{Scala2020, Neira2022}, we selected the test case as the Formation Flying L-band Aperture Synthesis (FFLAS) study. Its characteristics are described in Table II, where the relative position in the radial-transversal-normal (RTN) frame is reported. Moreover, Table III describes the physical characteristics of the array for the interferometer analysis\textcolor{black}{: the number of antennas per satellite ($N_{el}$), the antenna spacing ($d\cdot \lambda$) and the mean diameter for each satellite ($D_{mean}$)}.

\begin{figure}[H]
\centering
\subfloat[]
{
\includegraphics[scale = 0.3]{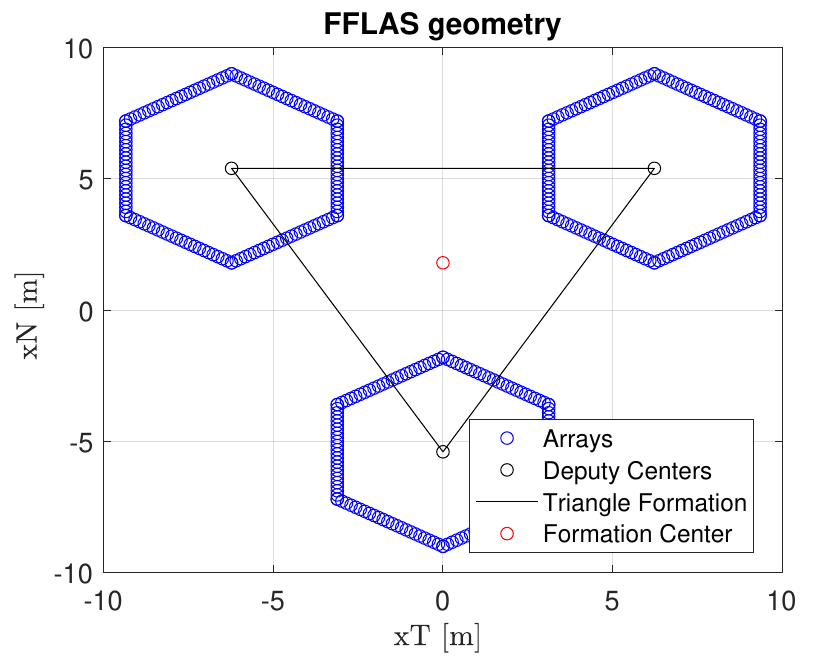}
         \label{fig:FFLAS_geometry}
}
\subfloat[]
{
 \includegraphics[scale = 0.3]{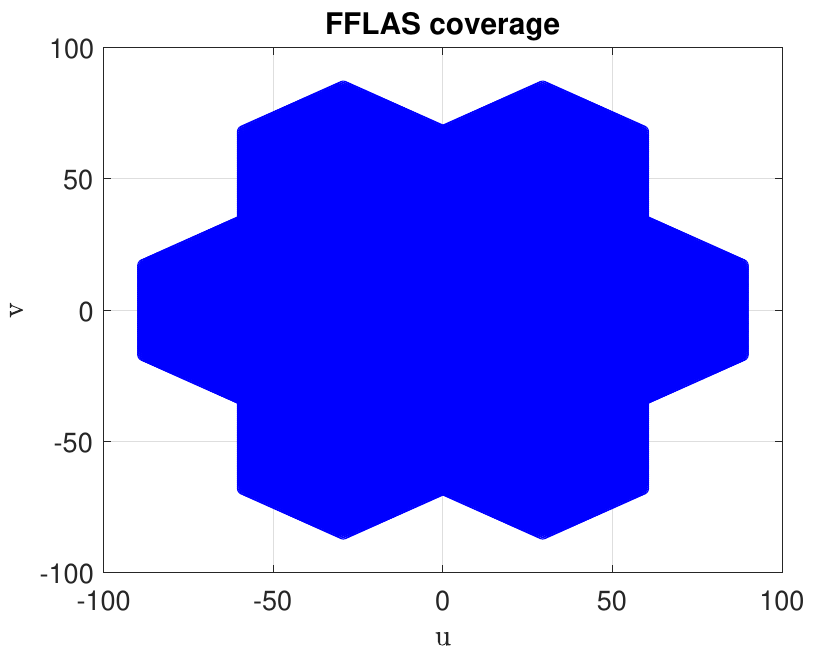}
       \label{fig:FFLAS_coverage}
       
}

\subfloat[]
{
 \includegraphics[scale = 0.3]{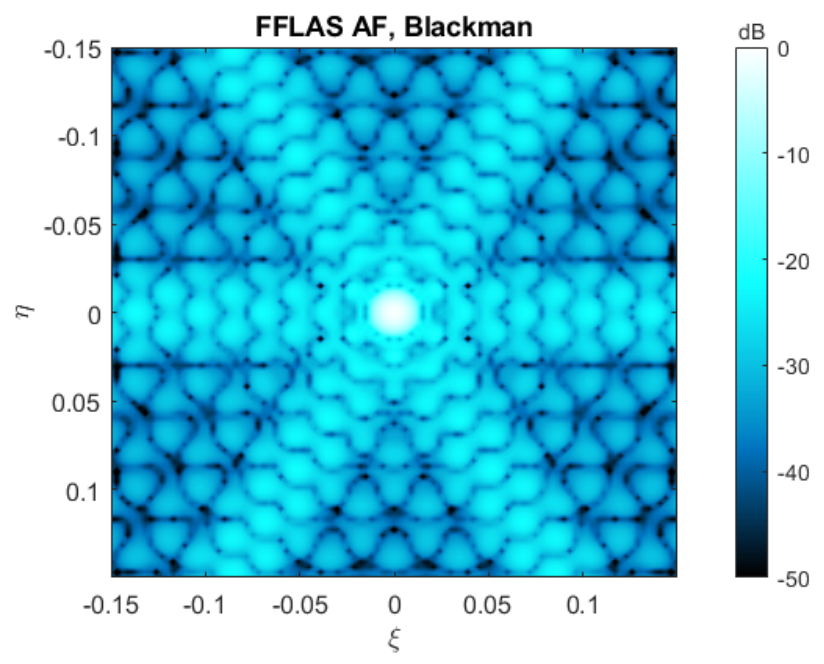}
      \label{fig:FFLAS_af}
}
   \caption{configuration (\ref{fig:FFLAS_geometry}), spatial frequency coverage (\ref{fig:FFLAS_coverage}) and array factor (\ref{fig:FFLAS_af}) for the FFLAS formation}
        \label{fig:FFLAS}
\end{figure}

Its geometry, spatial frequency coverage, and array factor are presented in Fig. \ref{fig:FFLAS}. The geometry is described in a Radial-Tangential-Nominal (RTN) frame, better explained in section \ref{subs:referencesystems}. 

\begin{table}[H]
\centering
\caption{Satellites' \textcolor{black}{center} position in the RTN frame for the FFLAS configuration.}
\begin{tabular}{|p{2 cm} | p{1.5 cm} |p{ 1.5 cm}| p{1.5 cm}|} 
\hline
 \rowcolor[HTML]{C0C0C0}
\textbf{Position [$m$]} & \textbf{Sat.1} & \textbf{Sat.2} & \textbf{Sat.3} \\ \hline \hline
$x_{R}$ & 0 & 0 & 0 \\ \hline
$x_{T}$ & 0 & 6.2350 & -6.2350  \\ \hline
$x_{N}$ & -5.3997 & 5.3997 & 5.3997\\ \hline
\end{tabular}
\label{tab:FFLAS_rtnframe}
\end{table}

\begin{table}[H]
\centering
\caption{Geometric characteristics for the FFLAS case.}
\begin{tabular}{|p{1 cm} | p{1.5 cm} |p{1.5 cm}| p{1.5 cm}  |} 
\hline
 \rowcolor[HTML]{C0C0C0}
\textbf{$N_{el}$} & \textbf{$d$} & \textbf{$\lambda$ $[cm]$} & \textbf{$D_{mean} $$[m]$} \\ \hline \hline
24 & 0.70724 & 21.209 & 7.2  \\ \hline
\end{tabular}
\label{tab:FFLAS_characteristics}
\end{table}

\section{Sensitivity analysis on Position error}
\label{sec:poserror}
This article aims to evaluate how an error in the relative position affects the payload performances, to what extent losses are present, and how the control system should act to restore the nominal condition. The relative position control is of utmost importance \citep{Scala2020, Neira2022}, and understanding how the payload behaves when some errors are present could help figure out what to expect and how to deal with them.

\begin{figure}[H]
\centering
\includegraphics[scale = 0.3]{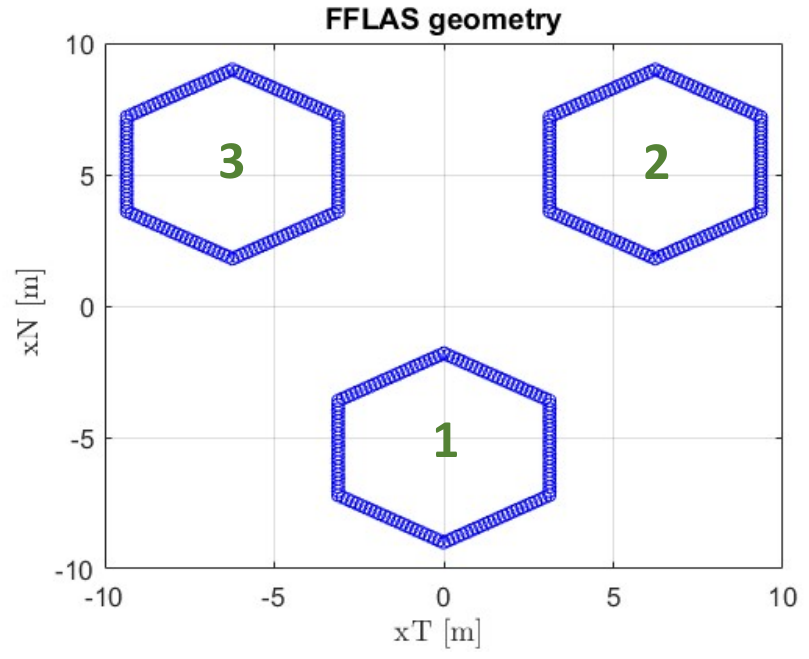}

   \caption{Satellites' order for the FFLAS configuration.}
        \label{fig:fflasnumber}
\end{figure}

This section presents a sensitivity analysis of the position error applied to FFLAS, focusing in particular on the \textcolor{black}{\textit{AF}} and \textcolor{black}{\textit{SLL}} matrices. The effect of a position error is visible also in an SFC plot but is less significant. For this reason, the coverages are not displayed. To better understand the following analyses, the satellites are named as 1, 2 and 3 as in Fig. \ref{fig:fflasnumber}. The focus is set on the RTN frame (see Section \ref{subs:referencesystems}), and only transversal and normal errors on the relative state of the formation are considered.

In general, errors can be subdivided into two categories: trajectory errors and attitude errors. The first type depends on position and velocity deviations, whereas the second one on the \textcolor{black}{pointing} stability of the satellite itself. The latter aspect is not investigated in this work. For this reason, the analysis is performed only on translation errors, while rotation errors are not taken into account. The focus is set only on position errors for simplicity, but since speed is the derivative of position, the results can be related to speed in future development.

\begin{figure}[H]
\centering

\includegraphics[scale = 0.3]{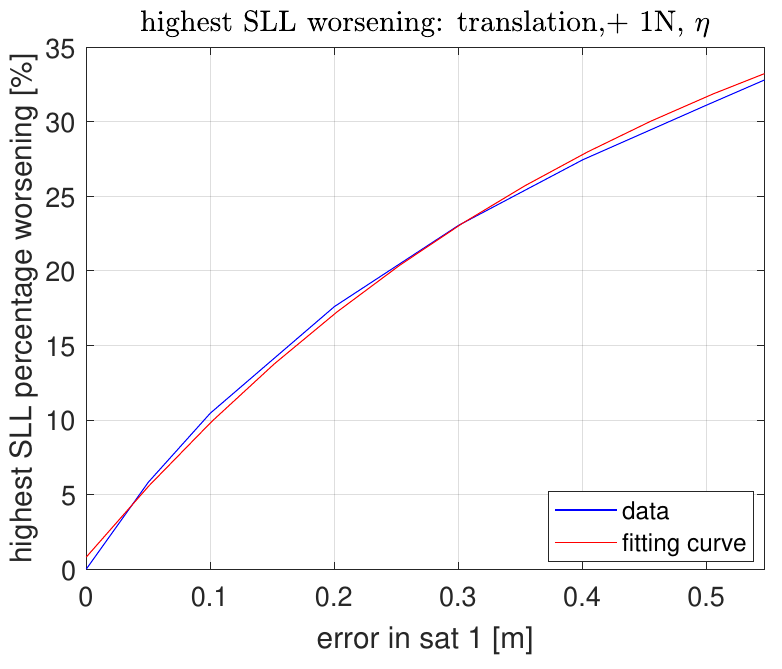}

   \caption{Fitting curve for the highest SLL worsening translation depending on the $\epsilon_{+1\textcolor{black}{N}}$ errors.}
        \label{fig:1y_datavsfitting}
\end{figure}

\begin{figure}[H]
\centering
\subfloat[]
{
 \includegraphics[scale = 0.3]{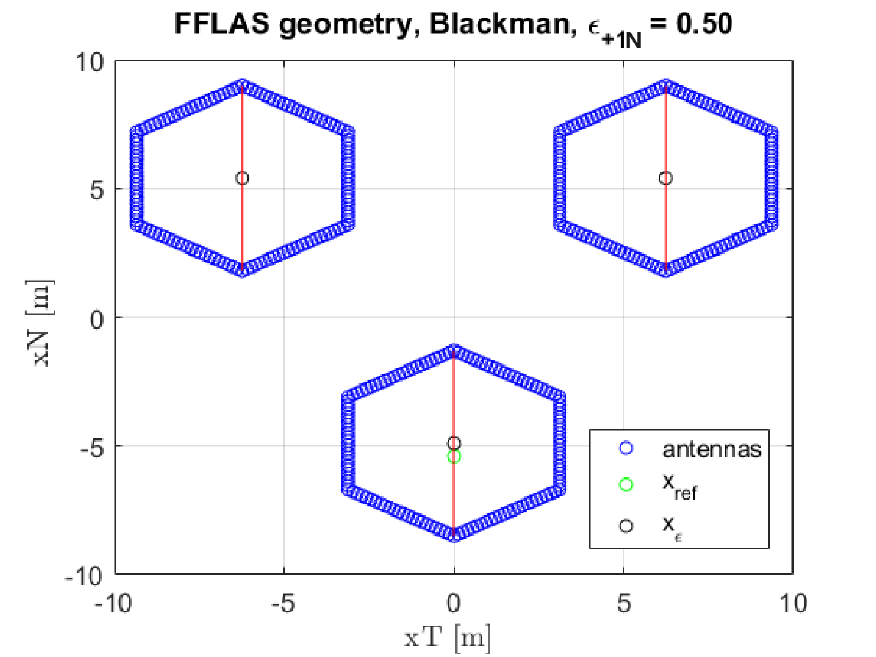}
     \label{fig:FFLAS_geom0.50}
}
\subfloat[]
{
 \includegraphics[scale = 0.3]{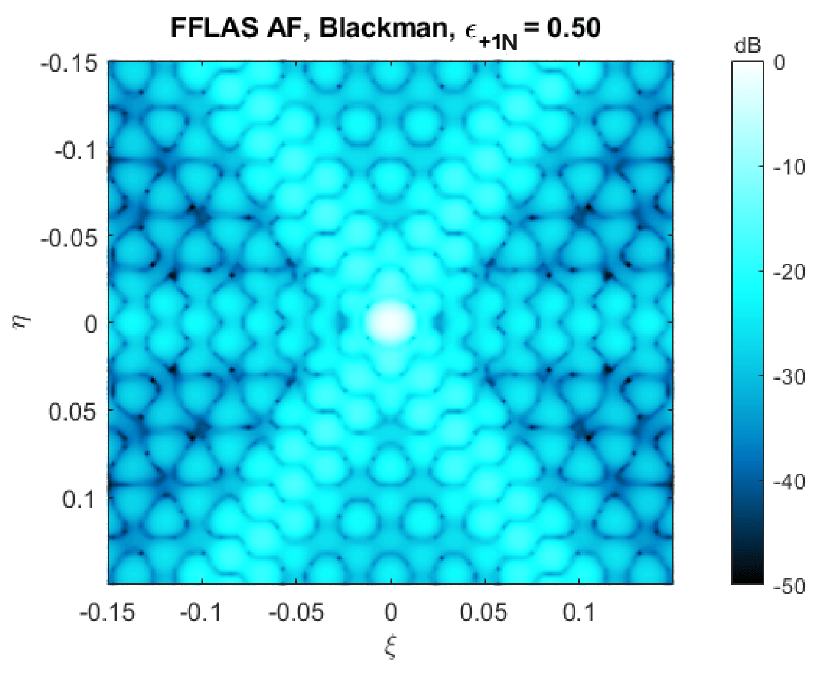}
     \label{fig:FFLAS_af0.50}
}

   \caption{Geometry and Array Factor for the $\epsilon_{+1\textcolor{black}{N}}$ = 0.5 $m$ case.}
   \label{fig:fflas1y0.5}
\end{figure}

To understand how a complex set of errors (i.e. multiple, multidirectional, and affecting more satellites) behaves, first, a simplified case is investigated. A single error in a single direction is evaluated, considering one satellite at a time \citep{Erbeia2022}. Each satellite is subjected to a single rigid translation. In particular, the error is defined as $\epsilon_{\alpha\beta}$ for satellite $\alpha$, with $\alpha$ = 1:3, in direction $\beta$, with $\beta$= $x_{N}$ or $x_{T}$. 
The notation for the satellite $\alpha$ is anticipated by a + or -- sign, depending on whether the deviation is along the positive or negative direction concerning the RTN frame (see Section \ref{subs:referencesystems}).
\textcolor{black}{The errors introduced in the simulation oscillate between 0 and 1 $m$. This decision is first made to understand how impactful a trajectory error can be on the payload itself, and which kind of worsening profile there could be (e.g. linear, exponential, logarithmic, etc, see Fig. \ref{fig:1y_datavsfitting}). Moreover, the satellites are considered with the nominal position as an initial starting point: satellites in such a close formation allow us to consider small errors. For instance, \cite{Scala2022} have demonstrated that the control accuracy for the FFLAS formation can be kept in $cm$-level, even considering navigation errors.}
The adopted procedure is the following:
\begin{enumerate}
    \item Compute the spatial frequency coverage (i.e. the Cartesian direction normalised by the wavelength) and the array factor (Eq. \ref{eq:AF}) having a position error as input;
    \item Evaluate the sidelobe level matrix in $\xi$ = 0 and identify the highest sidelobe level \textcolor{black}{(\textit{SLL})};
    \item Compute a percentage sidelobe level deviation \textcolor{black}{(here called \textit{Loss})} such as:
    \begin{equation}
        Loss = 100- \frac{Highest SLL (\epsilon)}{Highest SLL(\epsilon = 0)} \cdot 100  
    \end{equation}
     where \textcolor{black}{\textit{Highest SLL($\epsilon$ = 0)}} is the highest sidelobe in $\xi$ = 0 in the nominal condition, while \textcolor{black}{\textit{Highest SLL($\epsilon$)}} represents the same quantity, but when an error $\epsilon$ is present.
    This deviation is positive if there is an effective loss (so the highest sidelobe level is worse than the nominal one), otherwise, it is negative;
    \item repeat the procedure for different input errors;
    \item relate the percentage loss with the input error and compute a fitting curve with Matlab R2021b command Polyfit.
\end{enumerate}
To better understand the procedure, we first apply a positive translation along the normal direction $x_{N}$ for the first satellite (Satellite 1), which is therefore moving toward the center of the formation. Table \ref{tab:1yerror} describes how different values of the error $\epsilon_{+1\textcolor{black}{N}}$ influence the performance parameters. The highest value of the \textit{SLL} decreases progressively with larger error $\epsilon$, leading to a worsening of 43\%. The position in the ($\xi$, $\eta$) frame translates at around $\epsilon_{+1\textcolor{black}{N}}$ = 0.6 $m$. The collected data can be approximated by a Polyfit curve of degree 6 (Fig. \ref{fig:1y_datavsfitting}). The fitting coefficients are reported in Table \ref{tab:1ycoeff}. An example with an $\epsilon_{+1\textcolor{black}{N}}$ = 0.5 $m$ is given in Fig. \ref{fig:fflas1y0.5}.
\begin{table}[H]
\centering
\caption{Parameters of the $\epsilon_{+1\textcolor{black}{N}}$ normal error for the FFLAS case.}
\begin{tabular}{|p{1.3 cm} | p{3.6 cm} |p{ 1.05 cm}| p{1.3 cm} |} 
\hline
 \rowcolor[HTML]{C0C0C0}
\textbf{$\epsilon_{+1\textcolor{black}{N}}$ [$m$]}  & \textbf{\textit{Highest SLL} Value [$dB$]} & \textbf{Position} & \textbf{\textit{Loss} [$\%$]} \\ \hline \hline
0  & -20.3705 & 0.024 & 0 \\ \hline
0.05 & -19.1832  & 0.024 & 5.8285 \\ \hline
0.1  & -18.2391  & 0.024 & 10.4632 \\ \hline
0.2  & -16.7825  & 0.024 & 17.6137 \\ \hline
0.3  & -15.6731  & 0.024 & 23.0598 \\ \hline
0.4 & -14.7778  & 0.024 & 27.45498 \\ \hline
0.5  & -14.0292  & 0.024 & 31.1298 \\ \hline
0.6  & -13.2929   & 0.021 & 34.7444 \\ \hline
0.7 & -12.7474   & 0.021 & 37.4223 \\ \hline
0.8 & -12.2777  & 0.021 & 39.7280 \\ \hline
0.9 & -11.8692  & 0.021 & 41.7334 \\ \hline
1  & -11.5114  & 0.021 & 43.4899 \\ \hline
\end{tabular}
\label{tab:1yerror}

\end{table}

\begin{table}[H]
\centering
\caption{Coefficients for the fitting curve of Fig. \ref{fig:1y_datavsfitting} for the $\epsilon_{+1\textcolor{black}{N}}$ errors.}
\begin{tabular}{ |p{1.5 cm} |p{1.5  cm} | p{1.5  cm} | p{1.5  cm} |} 
\hline
 \rowcolor[HTML]{C0C0C0}
& \textbf{$x^{6}$} & \textbf{$x^{5}$} & \textbf{$x^{4}$}  \\ \hline \hline
& -0.2210 & 3.4350 & -20.8205  \\ \hline
 \rowcolor[HTML]{C0C0C0}
 \textbf{$x^{3}$} & \textbf{$x^{2}$}  & \textbf{$x$}  & \textbf{$-$}  \\ \hline
62.6261 & -102.3127 & 99.6770 & 0.8058  \\ \hline

\end{tabular}
\label{tab:1ycoeff}

\end{table}

Starting from the outcomes of the work presented in \cite{Neira2022}, a minimum inter-satellite distance of 12 $m$ has been identified for the FFLAS case. \textcolor{black}{If satellite 1 shifts towards the other two, it means that the formation goes from an equilateral triangle of 12.47$m$-side to an isosceles triangle of, as a minimum, 12 $m$-side. To satisfy this constraint, the maximum deviation for an $\epsilon_{+1\textcolor{black}{N}}$ error, $\epsilon_{1\textcolor{black}{N},max}$, could be 0.5463 $m$. This reasoning has been performed considering only the geometrical features (i.e., the relative distance) for the FFLAS formation.} This value is considered as the boundary value for the x-axis in Fig. \ref{fig:1y_datavsfitting}.  
Consider now an efficiency parameter $\beta$, defined as:
\begin{equation} \label{eq:beta}
    \beta = 1- \frac{fitting(\epsilon)}{fitting(\epsilon_{1y,max})},
\end{equation}
where \textcolor{black}{\textit{fitting}} represents the Polyfit fitting curve (Table \ref{tab:1ycoeff}). The efficiency parameter is set equal to 1 in the nominal case and 0 when $\epsilon_{1\textcolor{black}{N}}$ = $\epsilon_{1\textcolor{black}{N},max}$, and can further define the performance loss with respect to the nominal case. For instance, an $\epsilon_{+1\textcolor{black}{N}}$ of 0.023 $m$ leads to a percentage worsening of 3.0450, and a $\beta$ = 0.9084. The performance is therefore at 90.84$\%$ from the nominal case.

From the evaluation of a single error for each satellite, some considerations are derived \citep{Erbeia2022}:
\begin{itemize}
    \item a symmetrical error corresponds to a symmetrical graph. Therefore, an array factor output will be symmetric if the formation itself is symmetric, whereas asymmetric array factor plots are generated if the formation is no more symmetric. This consideration is also reflected in the formation centroid. As proof of this, FFLAS is also evaluated with an inter-satellite distance equal to the minimum possible, so at 12 $m$ (Fig. \ref{fig:FFLASminimum}). The plot is altered with respect to Fig. \ref{fig:FFLAS_af}, but it is still symmetric;
    \begin{figure}[H]
\centering

\includegraphics[scale = 0.3]{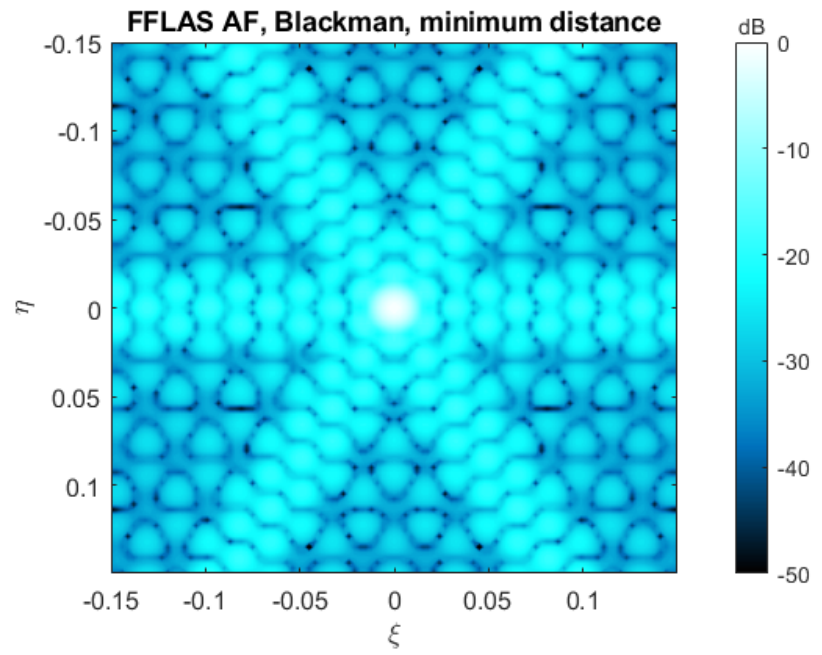}

   \caption{Array factor for the FFLAS configuration computed at the minimum allowable inter-satellite distance.}
        \label{fig:FFLASminimum}
\end{figure}
\item not only the plot is asymmetric if uneven errors are present, but it also exhibits preferential directions according to this error. That is why either sidelobe level improvements or worsenings, and differences in modulus are observed when comparing only the highest sidelobe levels.  Indeed, they are all evaluated for the $\xi$=0 cut, which can be misaligned from the preferential direction;
\item a mirrored error corresponds to a mirrored graph. Consider for instance Fig. \ref{fig:FFLAS_af0.50}: the plot is mirrored as the first satellite keeps an equal distance from satellite 2 and 3;
\item an approach between satellites leads to a more intense sidelobe level worsening than an estrangement. This is evident comparing Fig. \ref{fig:FFLAS_af0.50} with Fig. \ref{fig:af_1y_-0.5};
\begin{figure}[H]
\centering

 \includegraphics[scale = 0.3]{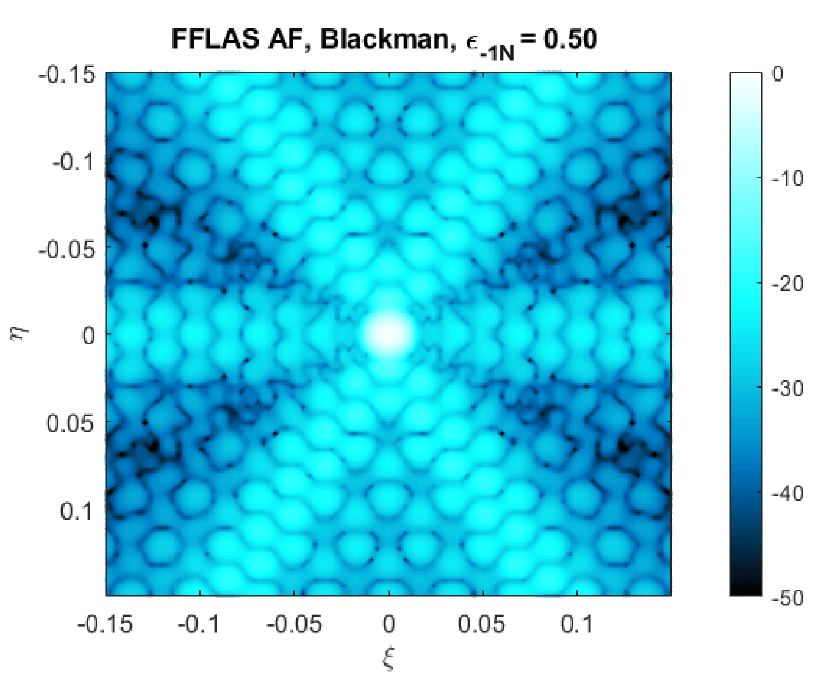}
   \caption{Array Factor for the $\epsilon_{-1\textcolor{black}{N}}$ = 0.5 $m$ case.}
   \label{fig:af_1y_-0.5}
\end{figure}

\item particularly for displacements along $x_{N}$, an approach between satellites leads to an approach of the highest sidelobe level position to the main lobe, and vice-versa.
\end{itemize}
There seems therefore to exist a relationship between the modulus and direction of an error, and the resultant array factor. \textcolor{black}{The effects of symmetry/ asymmetry highlighted in the previous considerations are influenced by the formation itself and by how errors propagate in one direction only. The correlations are more evident than in a realistic use case, in which multiple, multidirectional errors might be present. The aim of this process is indeed to bring out and recognise any key behaviours that the FFLAS formation shows in the presence of errors. Intuiting possible correlations with simplified errors can allow us to evaluate more complex scenarios. For example, it could be verified whether the \textit{AF} obtained from the presence of multiple errors can be decomposed into \textit{AFs} calculated via unidirectional errors, according to a principle of superposition of effects.} As a consequence \textcolor{black}{of the previous considerations}, we defined the following parameters for a better understanding of the problem. In particular \citep{Erbeia2022}:
\begin{itemize}
    \item main lobe: the area in which the main lobe and the first side lobe are included (black circle, Fig. \ref{fig:FFLAS_mainlobe}). It is defined as:
    \begin{itemize}
        \item Asymmetric (A) if a strong asymmetry is shown;
        \item Symmetric (S) if the main lobe remains more or less symmetrical;
        \item Symmetric/Asymmetric (S/A) if the main lobe is a hybrid between the two cases.
    \end{itemize}
    \item tail: the main six sidelobe forkings visible in Fig. \ref{fig:FFLAS_tails}. They are monitored from an \textcolor{black}{\textit{SLL}} point of view, focusing on the centres of the 'dots' making the tails. The positive signs mean that the tails show higher values than the nominal case, while the negative ones represent the opposite situation. In particular:
    \begin{itemize}
        \item   If the \textit{SLL} values are bigger than or equal to -16.2 $dB$, the tail is described by the parameter '\textbf{++}';
        \item if they are in majority between -16.2 $dB$ and -19.8 $dB$, the tail is described by the parameter '\textbf{+}';
        \item if they are in majority between -19.8 $dB$ and -22 $dB$, the tail is described by the parameter '$\mathbf{\sim}$';
        \item if they are in majority less than -22 $dB$ and the tail structure is more or less maintained, the tail is described by the parameter '\textbf{-};
        \item if they are in majority less than -22 $dB$ but the tail structure is hardly maintained, the tail is described by the parameter '\textbf{- -}'.
    \end{itemize}
    These values are selected as a qualitative threshold, to highlight the main deviation from the nominal, error-less case.

    \begin{figure}[H]
\centering
\subfloat[]
{
 \includegraphics[scale = 0.2]{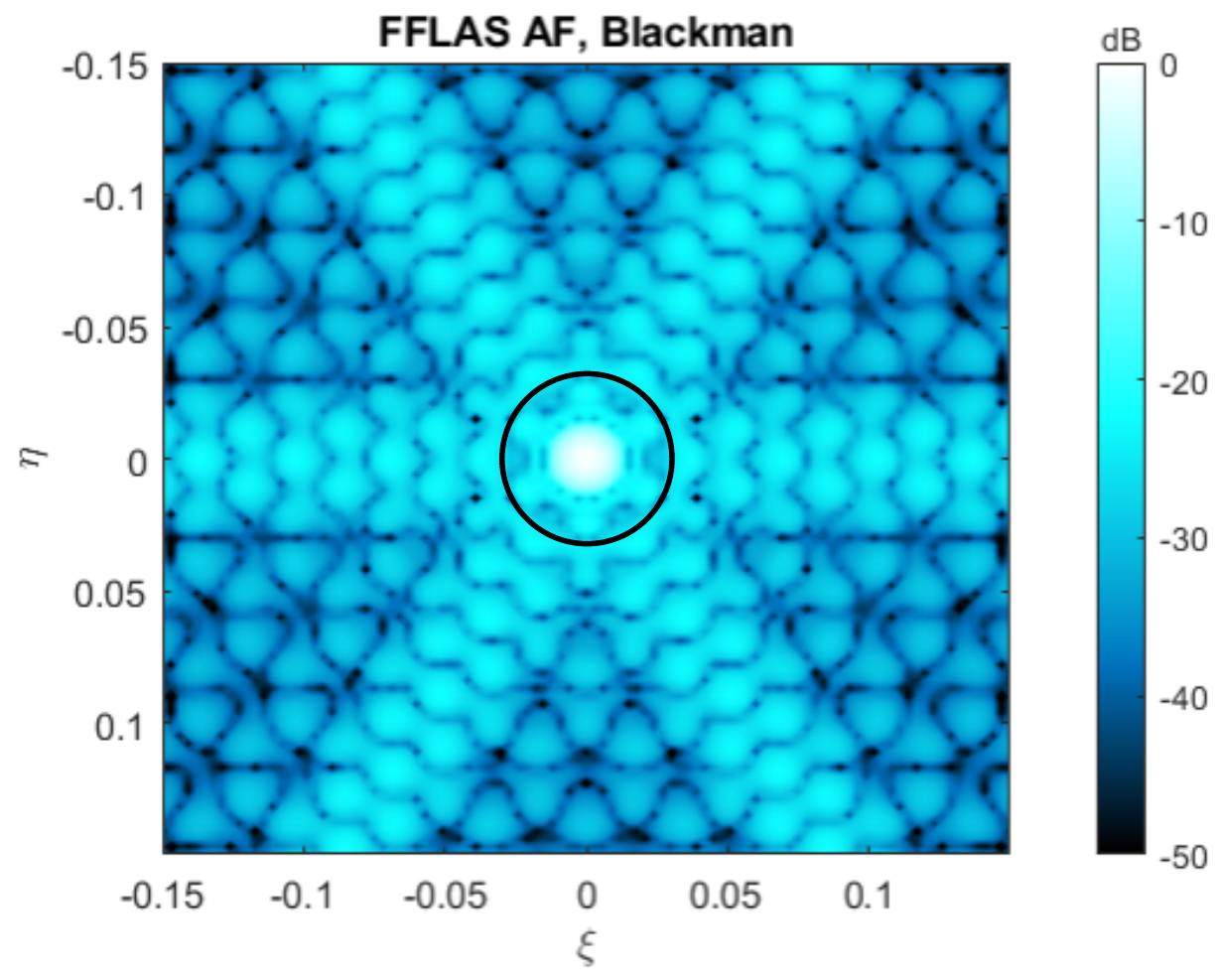}
     \label{fig:FFLAS_mainlobe}
}
\subfloat[]
{
 \includegraphics[scale = 0.2]{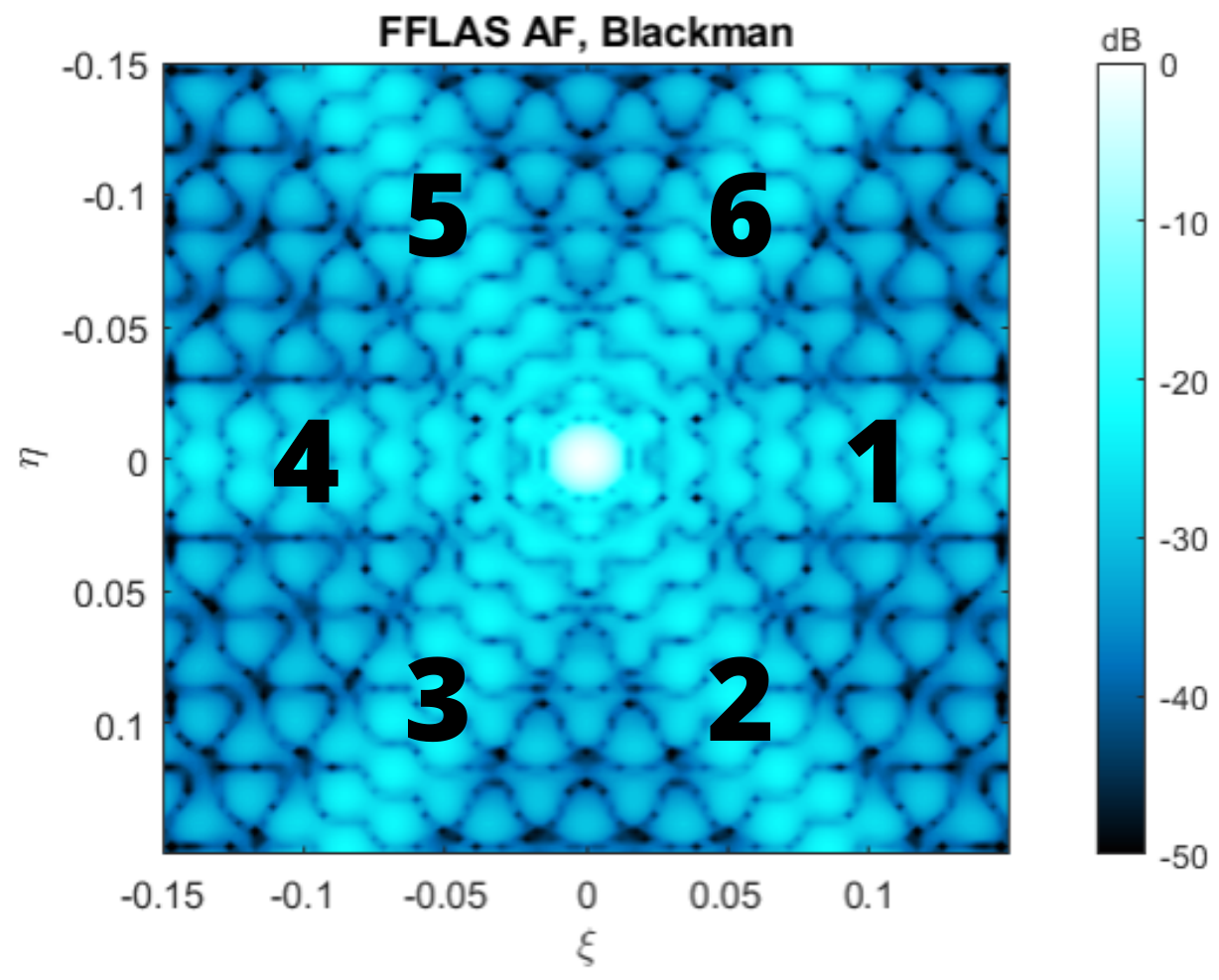}
     \label{fig:FFLAS_tails}
}

   \caption{Satellites' mainlobeThese  (\ref{fig:FFLAS_mainlobe}) and tails (\ref{fig:FFLAS_tails}) definition}
        
\end{figure}  

        \item R: the areas between the tails. The sides of the selected sector are indicated as a subscript for the \textcolor{black}{\textit{R}} letter, eventually underlined to mean a particular closeness. For example, $R_{\underline{1}2}$ defines the area between tails 1 and 2, near the first one. Some plots are affected by some low-\textcolor{black}{\textit{SLL}} oscillations. Because of the chromatic choice applied to the plots and because of the resemblance with a wave in the sea, these oscillations are here called blue waves. 
    The areas are monitored to see whether this 'blue wave' is more evident in the plot or not. 
    
    \item V or L: if the plot is the result of an approach between satellites or a departure respectively.
\end{itemize}

\textcolor{black}{These parameters have been applied considering the errors for the first and second satellites. $\epsilon_{3}$ and $\epsilon_{-1\textcolor{black}{T}}$ have not been computed again, as symmetric results are expected, as derived in \citep{Erbeia2022}.}
Considering for instance a comparison of a single error $\epsilon$ = 0.6 $m$, it derives that: 
\begin{table}[H]
\centering
\caption{Results for the $\epsilon$ = 0.6 comparison.}
\begin{tabular}{|p{1.2 cm} |p{0.6 cm} | p{0.7 cm} |p{0.6 cm}| p{0.6 cm} | p{0.5 cm} | p{0.5 cm} | p{0.5 cm}  |} 
\hline
   \rowcolor[HTML]{C0C0C0}
\textbf{Variable} & \textbf{$\epsilon_{+1\textcolor{black}{T}}$} & \textbf{$\epsilon_{-1\textcolor{black}{N}}$} & \textbf{$\epsilon_{+1\textcolor{black}{N}}$} & \textbf{$\epsilon_{-2\textcolor{black}{T}}$}  & \textbf{$\epsilon_{+2\textcolor{black}{T}}$}  & \textbf{$\epsilon_{-2\textcolor{black}{N}}$}   & \textbf{$\epsilon_{+2\textcolor{black}{N}}$} \\ \hline \hline
Main Lobe & S/A & A & S & A & A & A & S/A \\ \hline
Tail 1/4 & - & - & - & ++ & + & $\sim$ & $\sim$ \\ \hline
Tail 2/5 & $\sim$ & + & ++ & - & - - & - - & - - \\ \hline
Tail 3/6 & + & + & ++ & + & $\sim$  & ++ & + \\ \hline
Blue Wave & $R_{\underline{2}3}$, $R_{\underline{5}6}$ & 
$R_{tail 4}$, $R_{tail 1}$ & none & none & $R_{2\underline{3}}$, $R_{5\underline{6}}$ & $R_{\underline{1}2}$, $R_{\underline{4}5}$ & $R_{3\underline{4}}$, $R_{6\underline{1}}$ \\ \hline
V/L & L & L & V & V & L & V & L \\ \hline  
\end{tabular}
\label{tab:0.6comparison}
\end{table}

\begin{itemize}
    \item an error is present whether the plot shows some kind of asymmetry or the colour grid is altered than the nominal case;
    \item if tails 1/4 have worsened particularly, a $\epsilon_{-2\textcolor{black}{T}}$ or a $\epsilon_{+3\textcolor{black}{T}}$ error should be present. If they have worsened, but slightly less, then they should be referred to as a $\epsilon_{+2\textcolor{black}{T}}$ or a $\epsilon_{-3\textcolor{black}{T}}$ error. They are therefore not linked to the first satellite;
    \item if tails 2/5 have worsened particularly, a $\epsilon_{+1\textcolor{black}{N}}$ error should be present. If they have worsened, but slightly less, then they should be referred to as a $\epsilon_{-1\textcolor{black}{N}}$, a $\epsilon_{-2\textcolor{black}{T}}$ or a $\epsilon_{+3\textcolor{black}{T}}$ error;
    \item tails 3/6 should always have worsened when an error in the first satellite is present;
    \item an approach between satellites usually leads to a stronger sidelobe level worsening in the tails than a departure;
    \item if the approach is due to a $\epsilon_{-2\textcolor{black}{N}}$ or a $\epsilon_{-3\textcolor{black}{N}}$ error, some kind of blue waves should be observable; otherwise, they should not be present, and all the \textit{R} should appear more uniform;
    \item $R_{23}$ and $R_{56}$ show a blue wave if a departure is present. In particular, it should be nearer to tails 2 and 5 if a $\epsilon_{+1\textcolor{black}{T}}$ or a $\epsilon_{-3\textcolor{black}{T}}$ error is present, otherwise if it is due to a $\epsilon_{-1\textcolor{black}{T}}$ or a $\epsilon_{+2\textcolor{black}{T}}$ deviation;
    \item the blue wave should be present on both sides of tails 1 and 4 if a $\epsilon_{-1\textcolor{black}{N}}$ error is present;
    \item $R_{\underline{1}2}$ and $R_{\underline{4}5}$ show a blue wave if a $\epsilon_{-2\textcolor{black}{N}}$ or a $\epsilon_{+3\textcolor{black}{N}}$ error is present. However, for $\epsilon_{-2\textcolor{black}{N}}$ (so for an approach), it should be more visible;
    \item $R_{3\underline{4}}$ and $R_{6\underline{1}}$ should show a blue wave if a $\epsilon_{+2\textcolor{black}{N}}$ or a $\epsilon_{-3\textcolor{black}{N}}$ error is present. However, for $\epsilon_{+3\textcolor{black}{N}}$ (so for an approach), it should be more visible.
\end{itemize}
This analysis demonstrates that there exists an intrinsic link between the kind of position error, its modulus and its direction with the modelling output. 
This method applies many simplifications, as it considers only the highest sidelobe, the cut in $\xi$=0, and few collected data for the fitting curve. Above all, it is not an automated process, and a case-sensitive study is required for the position definition \citep{Erbeia2022}. The evaluation of multiple errors can therefore prove to be critical.

To consider several, concurrent errors, a different wide strategy is applied. This time, the error is introduced simultaneously on the whole formation, both in the $x_{T}$ and $x_{N}$ (Fig. \ref{fig:FFLAS_geometry}) directions, checking that the minimum inter-satellite distance is respected. The error is generated considering a random number between 0 and 1, multiplied by a constant so to have a final value between -0.4 and 0.4 $m$. From this new configuration, an Array Factor and its related sidelobes are computed (Eqs. \ref{eq:AF} and \ref{eq:SLL}), and a new quantity \textcolor{black}{\textit{P}} is defined, such as: 
\begin{equation} \label{eq:Pmatrix}
    P(\epsilon) = \bigg |\frac{SLL(\epsilon)}{SLL_{0}}-1 \bigg | \cdot 100
\end{equation}
where $SLL_{0}$ is the nominal sidelobe level case, while  $SLL(\epsilon)$ the one with the set of random errors. The absolute value is added as all kinds of deviations are intended as errors, as the final target is the nominal \textcolor{black}{\textit{SLL}} condition. This new parameter is called percentage matrix, as it expresses the deviation from the nominal case as a percentage output.

\begin{figure}[H]
\centering
\subfloat[]
{
\includegraphics[scale = 0.3]{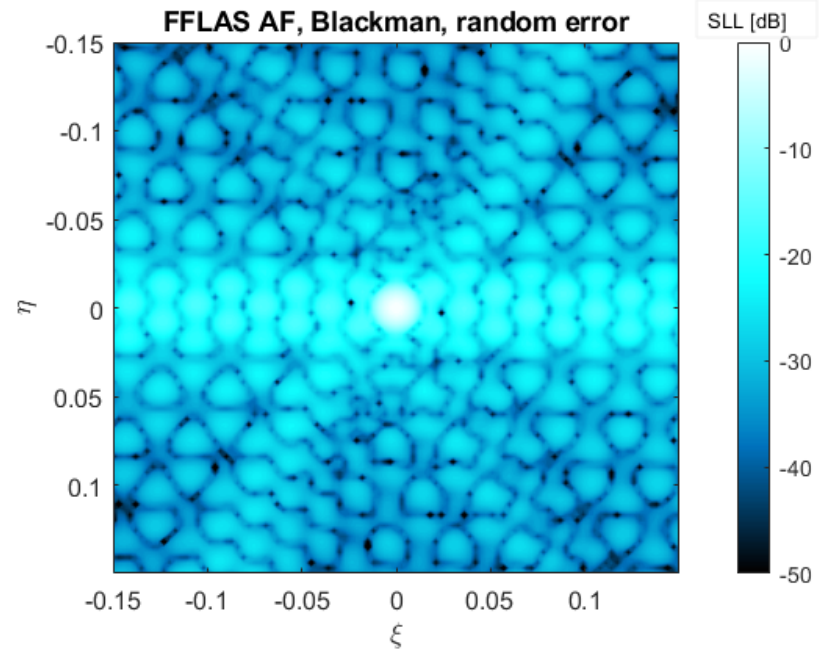}
 
}
\subfloat[]
{
\includegraphics[scale = 0.3]{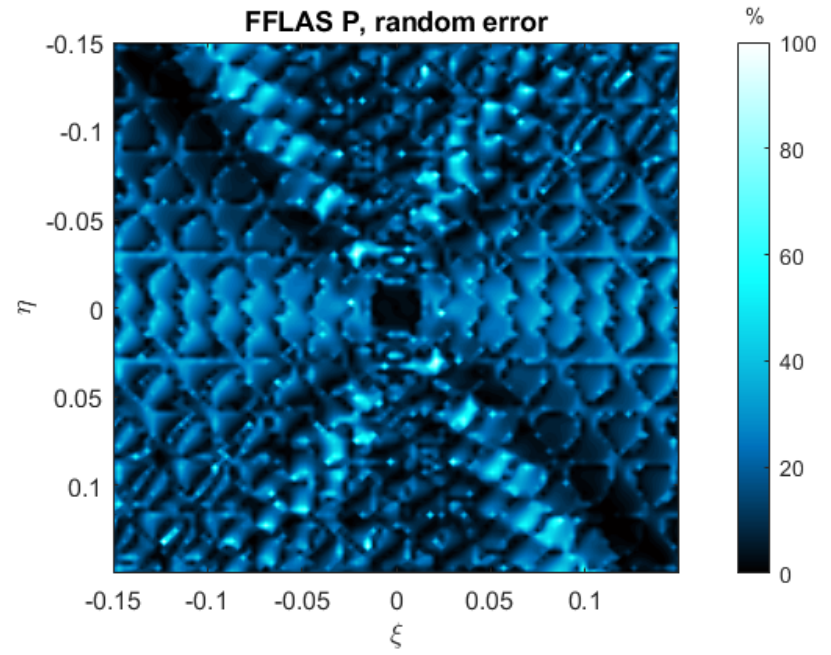}
   
}
   \caption{Array factor (left) and P matrix (right) for the random errors from Table \ref{tab:randomerr}}
        \label{fig:FFLAS_random}
\end{figure}

An example of this method is shown in Fig. \ref{fig:FFLAS_random}, where the errors from Table \ref{tab:randomerr} are applied.
\begin{table}[H]
\caption{Random input errors [$m$].}
\centering
\begin{tabular}{|p{0.8 cm} | p{1.1 cm} |p{ 1.1 cm} |p{0.8 cm} | p{0.9 cm} |p{1.1 cm} |} 
\hline
 \rowcolor[HTML]{C0C0C0} 
\textbf{$\epsilon_{1\textcolor{black}{T}}$} & \textbf{$\epsilon_{1\textcolor{black}{N}}$} & \textbf{$\epsilon_{2\textcolor{black}{T}}$} & \textbf{$\epsilon_{2\textcolor{black}{N}}$}  & \textbf{$\epsilon_{3\textcolor{black}{T}}$}  & \textbf{$\epsilon_{3\textcolor{black}{N}}$}  \\ \hline \hline
0.0703 & -0.2858 & -0.3582 & 0.1466 & 0.0868 & -0.2243\\ \hline

\end{tabular}
\label{tab:randomerr}

\end{table}

\begin{figure}[H]
\centering
\subfloat[]
{
\includegraphics[scale = 0.3]{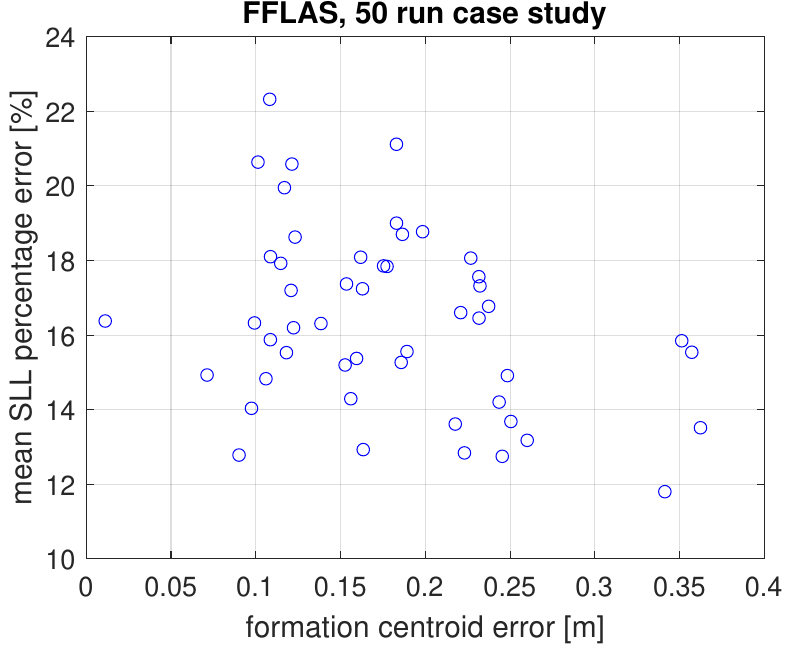}
}
\subfloat[]
{
\includegraphics[scale = 0.3]{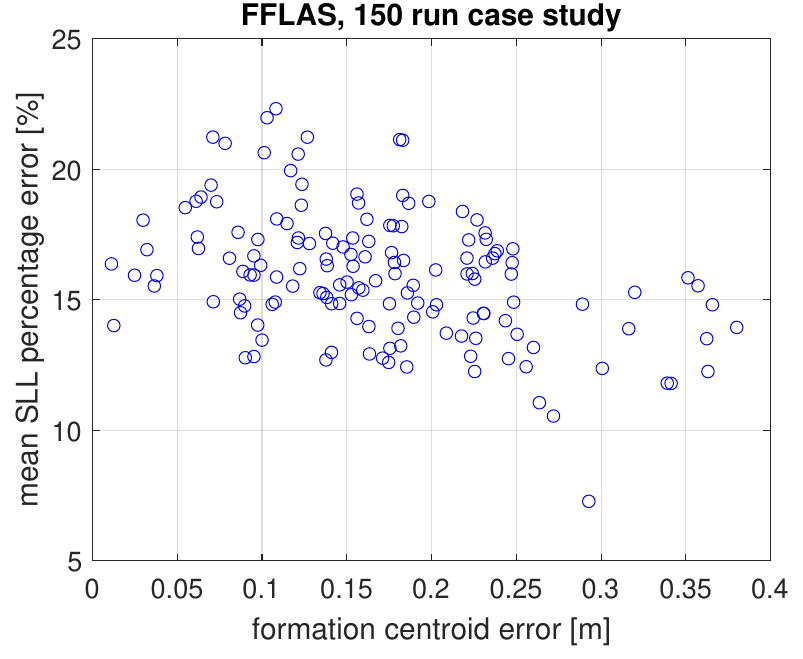}
  
}
   \caption{50 (left) and 150 (right) runs with random position errors for the FFLAS case}
        \label{fig:FFLAS_casestudies}
\end{figure}

The percentage matrix \textcolor{black}{\textit{P}} is obtained in an automated manner. Indeed, Table \ref{tab:randomerr} is just an example of possible random errors that can occur. It is, therefore, a 6x1 vector, being 6 as the number of errors and 1 as the only run that has been performed. This vector can be transformed into a 6xn matrix with an arbitrary number of runs n. A probabilistic approach can therefore be applied, considering each time a different column, so different initial conditions for computing the percentage matrix \textcolor{black}{\textit{P}}. From there, several considerations can be made, for instance, relating an average percentage (defined as the sum of all \textcolor{black}{\textit{P}} values divided by the total number of grid points) to the formation centroid deviation due to a random error. In this way, the centroid error is related to the average percentage and the number of runs, to follow the guidelines of a Montecarlo simulation, as in Fig. \ref{fig:FFLAS_casestudies}. \textcolor{black}{The formation centroid represents the center of gravity of a formation. Considering FFLAS in the TN plane and without errors, the coordinates of the formation centroid can be computed as $x^{formation}_{T}$ = $(x_{T1} + x_{T2}+ x_{T3})/{3}$ and $x^{formation}_{N}$ = $(x_{N1} + x_{N2}+ x_{N3})/{3}$, being $x_{T\alpha}$ and $x_{N\alpha}$ the coordinates of the satellites' center presented in Table \ref{tab:FFLAS_rtnframe}, with $\alpha = 1, 2, 3$. The formation centroid is represented by the red dot shown in Fig. \ref{fig:FFLAS_geometry}. The hypothesis of $x^{formation}_{R}$ always considered equal to 0 has also been introduced.
A displacement of one or more satellites in one or more directions leads to a displacement (so, to an error) of the centre of that/those satellites (see Fig. \ref{fig:FFLAS_geom0.50}) and, therefore, of the formation centroid. To keep consistency with the evaluations performed so far, the simplifications adopted, and with the \textit{P} matrix in Eq. \ref{eq:Pmatrix}, the displacement centroid error is considered by taking only its absolute value in the TN plane.}
The runs of the simulations consider a formation centroid error in the range 0.08-0.25 $m$ and a mean \textcolor{black}{\textit{SLL}} percentage error in 12$\%$-20$\%$ (see Fig. \ref{fig:FFLAS_casestudies}). Consequently, having a formation centroid error between $\pm$ 0.4 $m$, there is a high probability that the average loss will be between 12-20$\%$. Moreover, having a small formation centroid error does not mean having a small mean \textcolor{black}{\textit{SLL}} percentage error, and vice-versa: the two quantities are not proportional. \textcolor{black}{This non-correlation is due to the different nature of the parameters introduced. The mean \textit{SLL} percentage error is related to the relative position of the satellites within the formation: its deviations are influenced by considering errors applied to one satellite at a time. 
This quantity is influenced by the \textit{AF}, which depends on the SFC. For this reason, rigid drift of the whole formation in one direction would not lead to a significant change in the formation coverage itself (see section \ref{subs:FFLASunc} for some examples). The formation centroid error correlates instead the initial and the displaced position in the TN plane: in this case, a  drift of the whole formation in one direction would significantly influence the formation centroid error.} In a real case situation, the formation centroid error will not be probabilistic but dictated by the particular situation the satellite will be facing.

The method here described is more general with respect to the single-error one, as the entire sidelobe grid is considered. Moreover, it allows making different considerations, as the \textcolor{black}{\textit{P}} matrix could be analysed in several directions or related to multiple quantities. Finally, this technique can be automated, as shown in Fig. \ref{fig:FFLAS_casestudies}.

However, as the \textcolor{black}{\textit{P}} matrix is defined through an absolute value, it provides information only on the modulus of the deviation, not the direction (so if the sidelobe is improving or getting worse). A \textcolor{black}{\textit{P}} matrix without the absolute value could therefore be taken into account. The same limit is applied to the formation centroid deviation, as only the norm is considered. This is a limitation, as the same modulus could lead both to an approach or a departure between satellites, thus to different values for the same formation centroid error. For example, a $\epsilon_{+2\textcolor{black}{T}}$ deviation of 0.2378 $m$ leads to a formation centroid error of 0.0793 $m$, like a $\epsilon_{-2\textcolor{black}{T}}$ displacement of -0.2378 $m$. However, the average percentage errors resulting from $P_{2\textcolor{black}{T}}$ or $P_{-2\textcolor{black}{T}}$ differ, being respectively 12.8193$\%$ and 10.5926$\%$ \citep{Erbeia2022}.  
\textcolor{black}{It would be relevant to refine this approach and to apply it to the actual dynamics of the formation, introduced in the following section. In particular, the perturbations would introduce some multiple errors in multiple directions, which shall be studied simultaneously, as the \textit{P} matrix approach suggests.}


\section{FF control: Station keeping} \label{sec:control}
The implementation of the control is a critical aspect for formation flying, as station keeping is usually required with an accuracy of a few centimetres/ meters depending on the formation \citep{Neira2022}. For this reason, the dynamics of the satellite need to be described very accurately, to predict and implement an effective control action.

This study makes use of the linearised relative equations of motion for a formation of satellites, both in an unperturbed (the so-called Hill–Clohessy–Wiltshire equations, \cite{Alfriend2010, Clohessy1960, Hill1878}), and perturbed environment, in order to get to a state-space matrix \textit{A} and to have the dynamics described by $\mathbf{\dot{x}}$(t) = $A\mathbf{x}$(t). The perturbations are introduced in a linearised form, following the model proposed by \cite{Sabatini2008}, to include the $J_{2}$ and drag effects, that is, the two main effects in a LEO scenario. The simulations are run both in an uncontrolled and controlled environment, and the control is set through a Proportional-Integral-Derivative (PID) control \citep{Rocco2020}. The presented study is not a high-fidelity model, but it serves as a preliminary analysis of the relationship between payload modelling and relative control.

The focus is set on the FFLAS scenario. The dynamic of the formation is analysed also from a payload point of view. The configuration, the spatial frequency coverage, the array factor, and the \textit{P} matrix are displayed, and considerations about the mean sidelobe level deviations are performed. The unperturbed case is simulated for 2 orbital periods \textit{T}, while the perturbed ones are for 45\textit{T}. This latter value is selected as a compromise between having relevant results and limited time of observation, being $J_{2}$ and drag secular effects.

\subsection{Relative frame} \label{subs:referencesystems}
\begin{figure}[H]
\centering
\includegraphics[scale = 0.3]{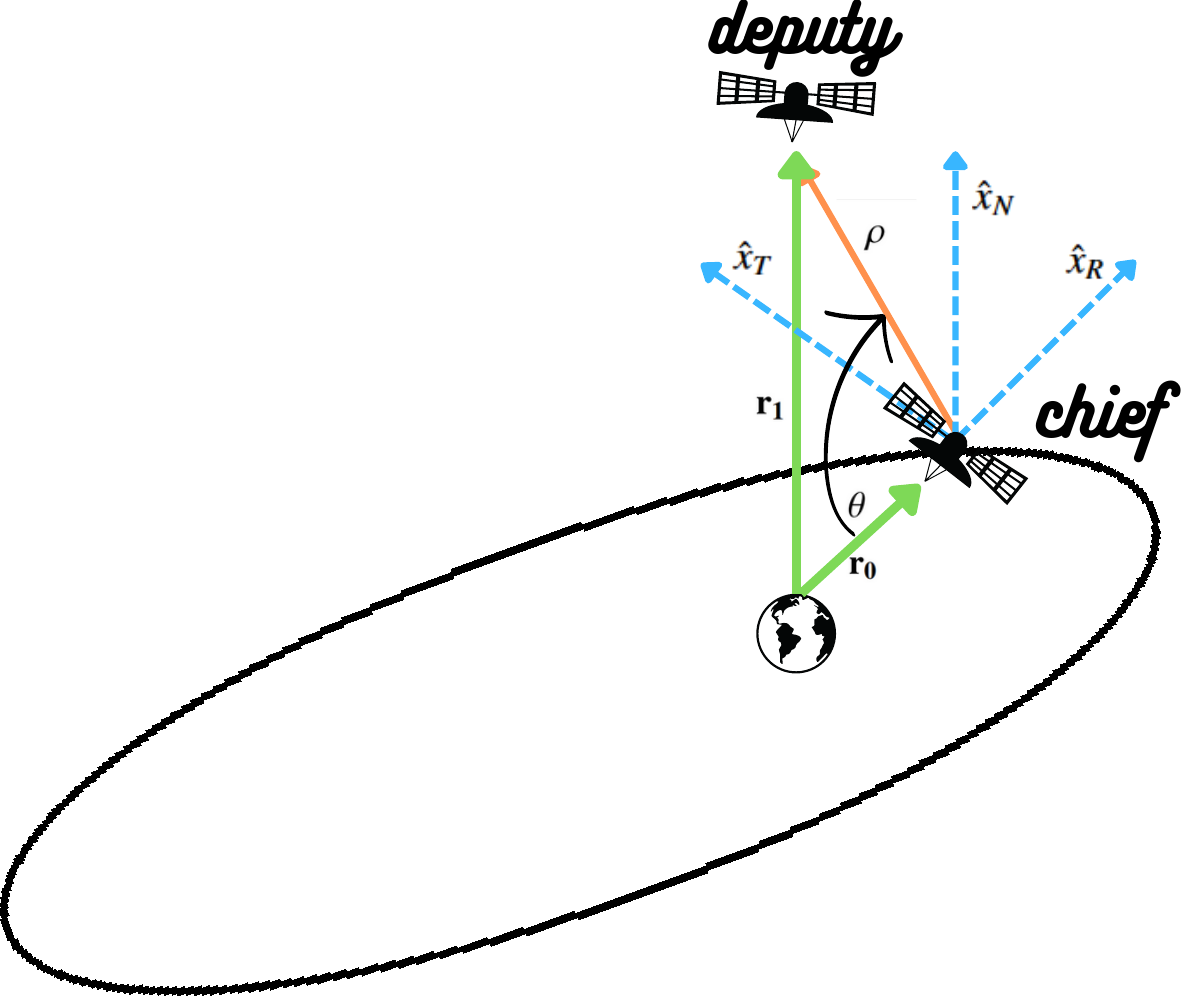}

   \caption{Representation of the RTN frame.}
        \label{fig:eh_frame new}
\end{figure}

The formation dynamics presented in this article is defined in a Radial, Transversal, Normal (RTN) frame. Considering a formation of two or more satellites, the main is the chief satellite, while all the others are called deputy satellites, and they orbit around the first one. This frame is a rotating coordinate system, with the origin centred in the chief. \textcolor{black}{The $x_{R}$-axis points outwards radially, $x_{N}$ points towards the direction of the (instantaneous) angular momentum, while $x_{T}$ completes the triad. The RT plane is therefore located in the chief's orbital plane, and the \textit{N}-axis is perpendicular to RT} (Fig. \ref{fig:eh_frame new}, \cite{Alfriend2010}).

\subsection{Formation under linearised $J_{2}$ and drag effect} \label{subs:j2+drag} 
The linear model including $J_{2}$ and Drag described in \cite{Sabatini2008} considers the two perturbations according to a principle of superimposition of effects. Some assumptions are applied, in particular: 

\begin{itemize}
    \item the initial orbit is considered circular or with a very low eccentricity;
    \item the inclination is kept constant;
    \item $\theta$ is considered as the sum of the true anomaly and of the argument of perigee, as function of time: $\theta$ = $n_{0}t$ + $\theta_{0}$, \textcolor{black}{with $\theta_{0}$= $\theta(t=0)$ and $n=\sqrt{\mu / r^{3}_{0}}$, being $r_{0}$ the initial radius}.
\end{itemize}

Unlike the $J_{2}$-only case, to solve the system of Equations, an auxiliary variable is introduced:
\begin{equation}
    x_{7} = -(\beta_{drag}-\alpha_{drag})\omega_{z},
\end{equation}

\textcolor{black}{where the subscript \textit{7} indicates that this auxiliary variable is the seventh component of the state space vector $\mathbf{x}$ = [\textcolor{black}{$x_{R}$}, \textcolor{black}{$x_{T}$}, \textcolor{black}{$x_{N}$}, $\textcolor{black}{\dot{x_{R}}}$, $\textcolor{black}{\dot{x_{T}}}$, $\textcolor{black}{\dot{x_{N}}}$, $x_{7}$]$^{T}$.} The parameters $\alpha_{drag}$ and $\beta_{drag}$ are the drag-associated parameter of the chief and of the deputy satellites, respectively, defined as:

 \begin{align*}
    \alpha_{drag} = \frac{1}{2} \frac{c_{D}A_{c}\rho_{atm} r_{c,0}}{M_{c}}, \\ 
    \beta_{drag} = \frac{1}{2} \frac{c_{D}A_{d}\rho_{atm} r_{d,0}}{M_{d}}, \\
    \end{align*}
    
where $c_{D}$ is the drag coefficient (equal to 2.2 in LEO, \cite{Sabatini2008}), while $M_{c,d}$, $A_{c,d}$ and $r_{c,d}$ represent the cross surfaces, the masses and the initial radii of chief and deputy. $\rho_{atm}$ is the density.
$\omega_{z}$ is the third component of the angular velocity $\omega$, defined as:
\begin{equation} \label{eq:omega_J2_drag}
    \mathbf{\omega}  = \begin{bmatrix} 
    \omega_{x} \\ \omega_{y} \\ \omega_{z}
    \end{bmatrix} =  \begin{bmatrix}
    \Dot{\gamma} \\ 0 \\ \frac{h}{r^{2}}
    \end{bmatrix},
\end{equation}
where \textcolor{black}{\textit{h}} is the orbit angular momentum (equal to $\sqrt{r\mu(1+e\cos{\theta})}$ for an elliptic orbit or simply $\sqrt{r\mu}$ for a circular one, \cite{Curtis2014}), and $\Dot{\gamma}$ = $\frac{r}{h} f_{h}$, being $f_{h}$= $-\frac{3}{2}J_{2}\mu\frac{R^{2}_{e}}{r^{4}} \sin({\theta})\sin({2i})$ the disturbing acceleration in the cross-track direction.

The state-space matrix is then augmented to a 7x7 matrix, and \textcolor{black}{$\mathbf{x}$ = [\textcolor{black}{$x_{R}$}, \textcolor{black}{$x_{T}$}, \textcolor{black}{$x_{N}$}, $\textcolor{black}{\dot{x_{R}}}$, $\textcolor{black}{\dot{x_{T}}}$, $\textcolor{black}{\dot{x_{N}}}$, $x_{7}$]$^{T}$} is the state vector.
The orbit radius is defined as:
\begin{equation}
    r=(r_{0}+X_{J_{2}})e^{-2\alpha_{drag}\theta}
\end{equation}

with
\begin{equation}
  X_{J_{2}} =
 \frac{3J_{2}R^{2}_{e}}{2r_{0}} \bigg [ \frac{1}{3}\sin^{2}({i})\cos^{2}({\theta}) + \frac{1}{3}\sin^{2}({i})-1 + \bigg ( 1- \frac{2}{3}\sin^{2}({i}) \bigg) \cos({\theta}) \bigg],   
\end{equation}
To compute \textcolor{black}{$x_{7}$}, $\dot{h}$ is needed. It is defined as $rf_{\theta}$, with $f_{\theta}$ = $f_{\theta, J_{2}}$ + $f_{\theta, drag}$:

\begin{equation}
 f_{\theta, J_{2}} = -\frac{3}{2}J_{2}\mu\frac{R^{2}_{e}}{r^{4}} \sin({2\theta})\sin^{2}({i}),  
\end{equation}
\begin{equation}
 f_{\theta, drag} = \frac{1}{2} \frac{\rho_{atm} c_{D}A_{d}\mu}{M_{d} r_{0}}.    
\end{equation}

The state matrix \textit{A} can now be defined as \citep{Sabatini2008}:

\begin{equation} \label{eq:A_J2+drag}
    A = \begin{bmatrix}
0 & 0 & 0 & 1 & 0 & 0 & 0 \\
0 & 0 & 0 & 0 & 1 & 0 & 0 \\
0 & 0 & 0 & 0 & 0 & 1 & 0 \\
a_{41} & a_{42} & a_{43} & a_{44} & 2\omega_{z} & 0 & a_{47}\\
a_{51} & a_{52} & a_{53} & -2\omega_{z} & a_{55} & 2\omega_{x} & a_{57}\\
a_{61} & a_{62} & a_{63} & 0 & -2\omega_{x} & a_{66} & 0 \\
0 & 0 & 0 & 0 & 0 & 0 & a_{77}\\
\end{bmatrix},
\end{equation}

where:
    \begin{align*}
    K &= \frac{6J_{2}\mu R^{2}_{e}}{r^{5}}, \\
         a_{41} &= \omega^{2}_{z} +2\frac{\mu}{r^{3}} + K(1-3\sin^{2}({i})\sin^{2}({\theta})), \\
         a_{42} &= \Dot{\omega_{z}} + K(\sin^{2}({i})\sin({2\theta})) + \beta_{drag} \omega^{2}_{z}, \\
        a_{43} &= -\omega_{x}\omega_{z} + K(\sin({2i})\sin({\theta})), \\
        a_{44} &= -\beta_{drag} \omega_{z}, \\
        a_{47} &= \dot{r}, \\
        a_{51} &= -\Dot{\omega_{z}}+K(\sin^{2}({i})\sin({2\theta})) -\beta_{drag} \omega^{2}_{z}, \\
        a_{52} &= \omega^{2}_{x}+\omega^{2}_{z} - \frac{\mu}{r^{3}} + K \bigg [ -\frac{1}{4} + \sin^{2}({i}) \bigg ( \frac{7}{4} \sin^{2}({\theta})-\frac{1}{2} \bigg ) \bigg], \\
        \end{align*}
        \begin{align*}
            a_{53} &= \Dot{\omega_{x}} + K \bigg ( -\frac{1}{4} \sin({2i}) \cos({\theta}) \bigg) + \beta_{drag} \omega_{x} \omega_{z}, \\
        a_{55} &= -\beta_{drag} \omega_{z}, \\
         a_{57} &= r\omega_{z}, \\
         a_{61} &= -\omega_{x}\omega_{z} + K(\sin({2i}) \sin({\theta})), \\
         a_{62} &= -\Dot{\omega_{x}} + K \bigg (-\frac{1}{4} \sin({2i}) \cos({\theta}) \bigg ) -\beta_{drag}\omega_{x}\omega_{z}, \\
        a_{63} &= \omega^{2}_{x} - \frac{\mu}{r^{3}} + K \bigg [-\frac{3}{4} + \sin^{2}({i}) \bigg (\frac{5}{4} \sin^{2}({\theta}) + \frac{1}{2} \bigg ) \bigg ], \\
        a_{66} &= -\beta_{drag}\omega_{z}, \\
        a_{77} &= -(\beta_{drag}-\alpha_{drag})\dot{\omega}_{z}.
    \end{align*}
    
A clarification needs to be made: the term $a_{45}$ is not equal to 0, as shown in \cite{Sabatini2008}, but is equal to 2$\omega_{z}$, as for the $J_{2}$ case.
The authors \cite{Sabatini2008} have verified that this linearised model matches the performances of already existing ones. For this reason, the model is considered reliable and effective enough to be used in a simulation.

\subsection{Proportional, integral, derivative controller} \label{subs:PIDcontrol}
To keep the satellites in the desired location, a control action \textit{\textbf{u}(t)} must be implemented. In this case, the state-space dynamics is described by $\mathbf{\dot{x}}(t)$ = $A\mathbf{x}(t)$ + \textcolor{black}{$B$}$\mathbf{u}(t)$.

For a preliminary assessment of the control performances and the link between the control and the error on the payload, a PID controller is considered.
Consider $\mathbf{x_{ref}}$ as the desired state vector, while $\mathbf{x}(t)$ the actual state vector at each instant of time. Then, an error vector $\mathbf{e}(t)$ can be defined as $\mathbf{e}(t)$ = $\mathbf{x}(t)$ - $\mathbf{x_{ref}}$, also called the Steady State (StS) error. 
\textcolor{black}{As a first, simplified approach, each satellite has a $\mathbf{x_{ref}}$ = [\textit{$x_{R}$; $x_{T}$; $x_{N}$; 0; 0; 0}] if only $J_{2}$ effect is present, with $x_{R}$, $x_{T}$, $x_{N}$ the satellites' position presented in Table \ref{tab:FFLAS_rtnframe}. In case of a $J_{2}$ + Drag scenario, $\mathbf{x_{ref}}$ = [\textit{$x_{R}$; $x_{T}$; $x_{N}$; 0; 0; 0; $x_{7}$}].}

A PID controller is defined as:
\begin{equation} \label{eq:PID}
    \mathbf{u}(t) = K_{p}\mathbf{e}(t) + K_{d}\mathbf{\dot{e}}(t) + K_{i}\int^{t_{f}}_{t_{0}} \mathbf{e}(t) dt,
\end{equation}

being $K_{p}$, $K_{d}$ and $K_{i}$ the proportional, derivative and integral  constant, respectively.

The situation described here can be seen as a system step response. For an analogy, there exist three main parameters that describe the output performance \citep{Rocco2020}:
\begin{itemize}
    \item the \textbf{rise time}, i.e. the time the response takes to get from 10$\%$ to 90$\%$ of the steady-state value;
    \item the \textbf{settling time}, i.e. the time needed to stay below 2$\%$ of $x_{0}$-$x_{ref}$;
    \item the \textbf{overshoot}, i.e. the maximum response range compared to the steady-state value, corresponding to the amplitude of the first peak.

\end{itemize}

The PID constants influence these three parameters (Table \ref{tab:PIDconstants_influence}, \cite{PID2017}). An accurate selection of $K_{p}$,  $K_{i}$ and $K_{d}$ helps therefore reaching the desired performance.
The control is implemented in a feedback configuration. 

\begin{table}[H]
\centering
\caption{Influence of the PID constants on an impulse response \citep{PID2017}.}
\begin{tabular}{ | p{1.3 cm} |p{ 1.2 cm}| p{1.2 cm} | p{1.4 cm}| p{1.1 cm}|} 
\hline
 \rowcolor[HTML]{C0C0C0}
 \textbf{Constant} & \textbf{Rise Time} & \textbf{Settling Time} &  \textbf{Overshoot} &  \textbf{StS Error}  \\ \hline \hline
$K_{p}$ & decrease & small change & increase & decrease  \\ \hline
$K_{i}$ & decrease & increase & increase & decrease \\ \hline
$K_{d}$ & small change & decrease & decrease & no change \\ \hline
\end{tabular}
\label{tab:PIDconstants_influence}
\end{table}

\subsection{Case study: FFLAS} \label{subs:FFLAS}
All the evaluations previously expressed are applied to the FFLAS scenario. The following considerations are taken into account:
\begin{itemize}
    \item the geometry characteristics are the ones set in Table \ref{tab:FFLAS_characteristics}, and the three satellites are set at an initial distance of 12.47 $m$;
    \item the three satellites are all considered as deputy, and a sort of fictitious chief satellite is considered in the origin of the RTN frame. For this reason, $\alpha_{drag}$ is set to 0;
    \item the fictitious chief orbits in a Sun-synchronous circular orbit, with a mean altitude $h_{0}$ = 774.85 $km$ and inclination \textit{i} = 98.44° \citep{Neira2022};
    \item the atmosphere is considered constant, such as $\rho_{atm}$ = 1.8102e$^{-14}$ $kg/m^{3}$. This value is taken considering an altitude of 774.85 $km$ and an exponential atmospheric model as described in \cite{Wertz1978};
    \item for $\beta_{drag}$ computation, $r_{d}$ is set equal to $r_{0}$;
    \item the payload characteristics are set as $M_{d}$ = 1600 $kg$ and $A_{d}$ = 3 $m^{2}$ \citep{Neira2022}.
\end{itemize}
 
\subsubsection{Uncontrolled case} \label{subs:FFLASunc}
\begin{figure}[H]
\centering
\subfloat[]
{
\includegraphics[scale = 0.3]{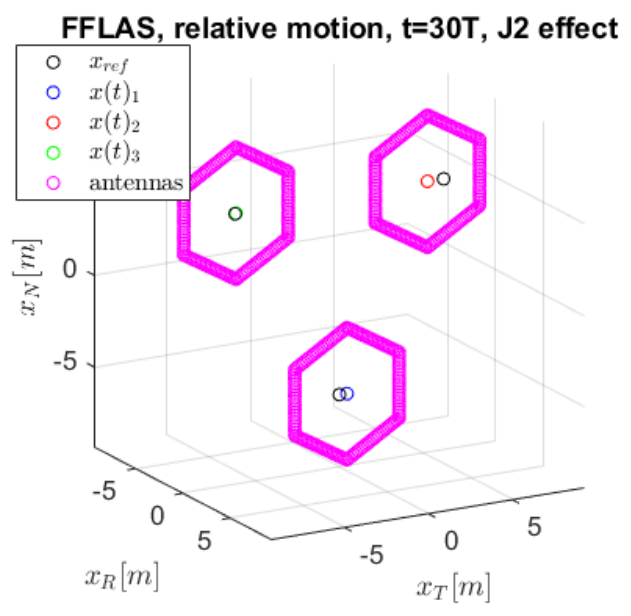}
}
\subfloat[]
{
 \includegraphics[scale = 0.3]{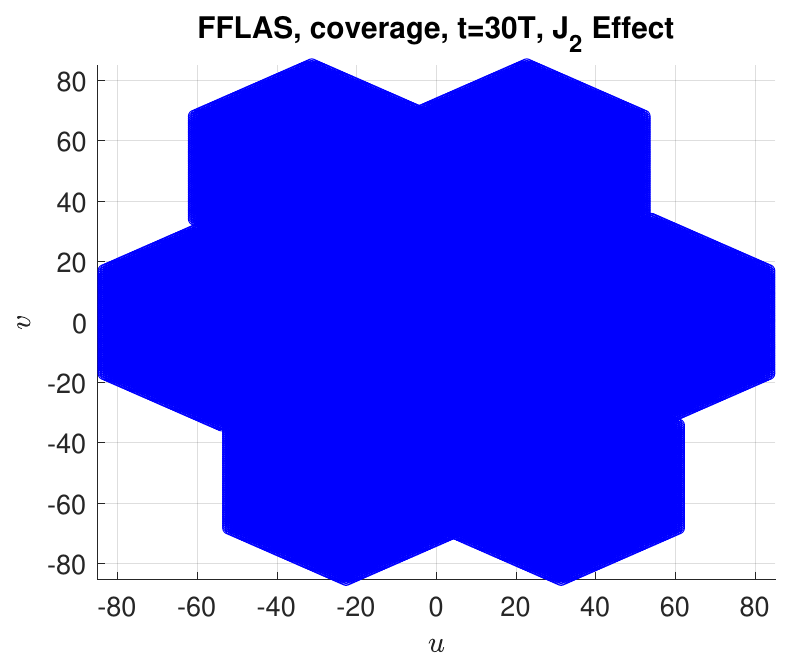}
     \label{fig:FFLAS_cov_30T_j2_nocontrol}
}
   
\subfloat[]
{
 \includegraphics[scale = 0.3]{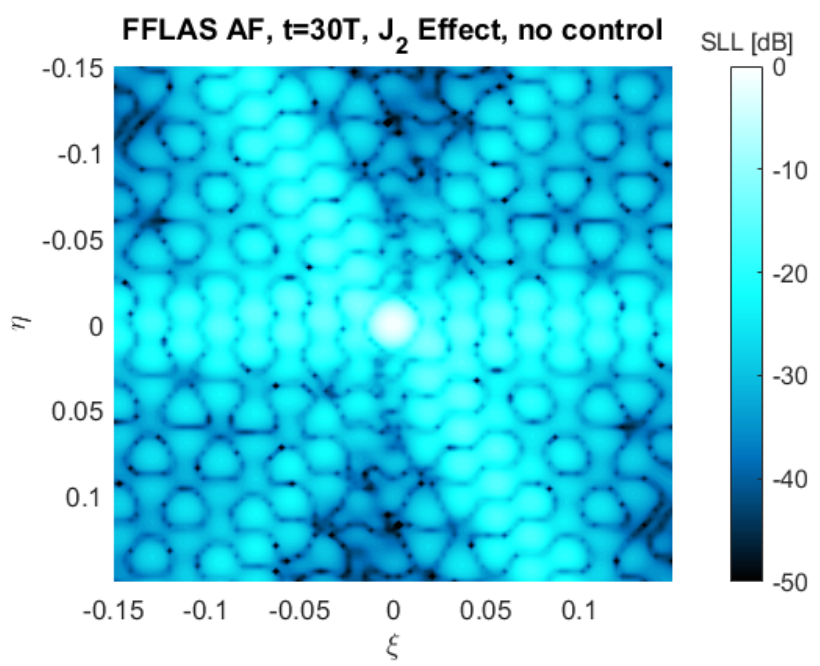}
}
\subfloat[]
{
\includegraphics[scale = 0.3]{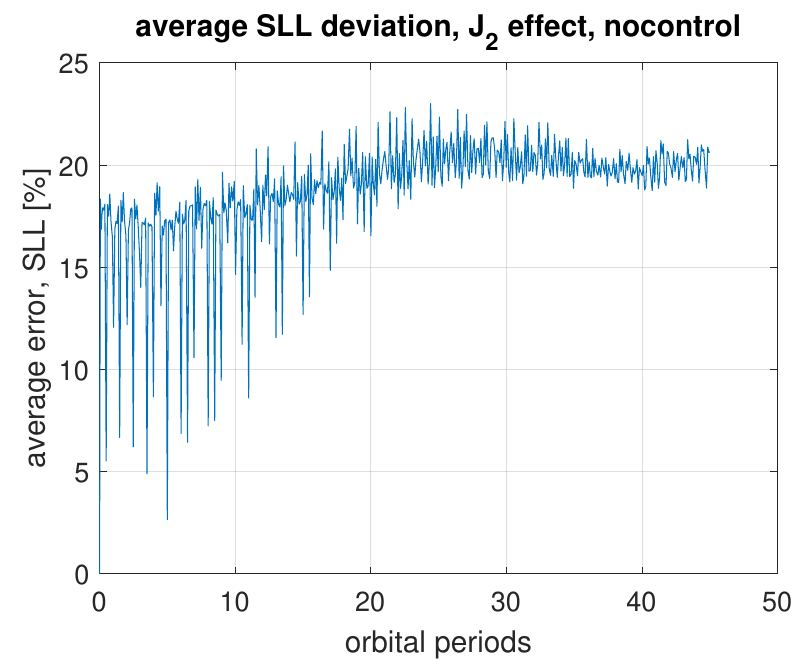}
\label{fig:avSLL_j2_period}
}
   \caption{AC, SFC, AF  at t=30T, and average P values for the FFLAS case under $J_{2}$ effect. No control.}
        \label{fig:FFLAS_payload_J2}
\end{figure}

\begin{figure}[H]
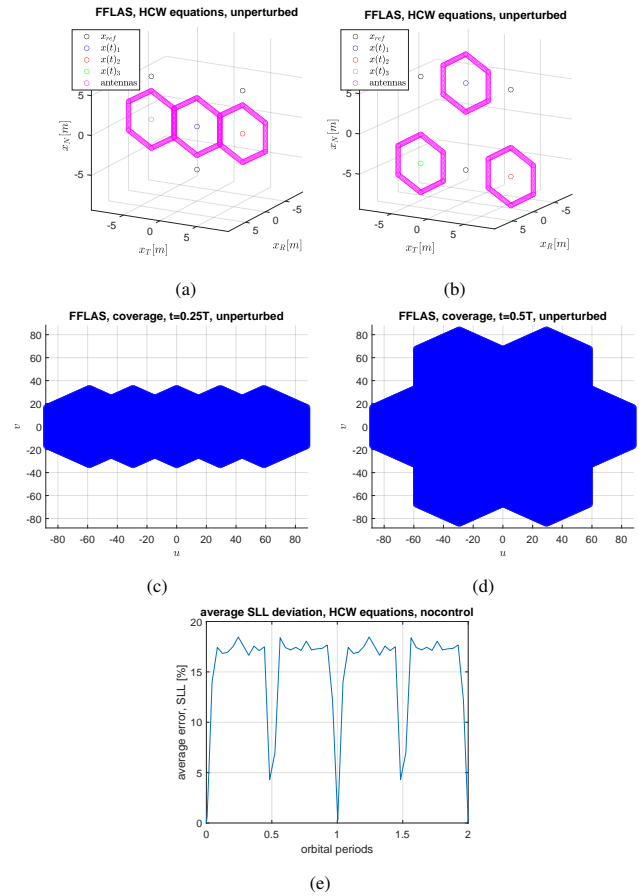

    \centering
    \subfloat[]{\includegraphics[scale = 0.3]{Images/FFLAS_geom_0.25T_nocontrol.pdf}
        }   
    \subfloat[]{\includegraphics[scale = 0.3]{Images/FFLAS_geom_0.5T_nocontrol.pdf}
        }
   
    \subfloat[]
    { \includegraphics[scale = 0.3]{Images/FFLAS_cov_0.25T_nocontrol.pdf}
        \label{fig:cov_0.25T_nocontrol}}
    \subfloat[]
    { \includegraphics[scale = 0.3]{Images/FFLAS_cov_0.5T_nocontrol.pdf}
         \label{fig:cov_0.5T_nocontrol}}

    \subfloat[]
    {\includegraphics[scale = 0.3]{Images/averageP_nocontrol.pdf}
         \label{fig:avSLL_unperturbed_period}}
     
    \caption{AC and SFC respectively t=0.25T (left) and t=0.5T (right), and average P values for the FFLAS case in an unperturbed motion. No control.}
    \label{fig:FFLAS_payload_unperturbed}
\end{figure}

\begin{figure}[H]
\centering
\subfloat[]
{
 \includegraphics[scale = 0.35]{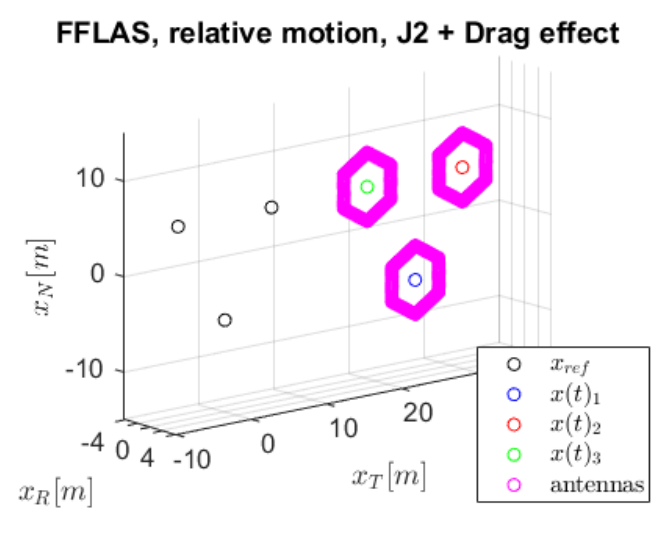}
}
\subfloat[]
{
\includegraphics[scale = 0.3]{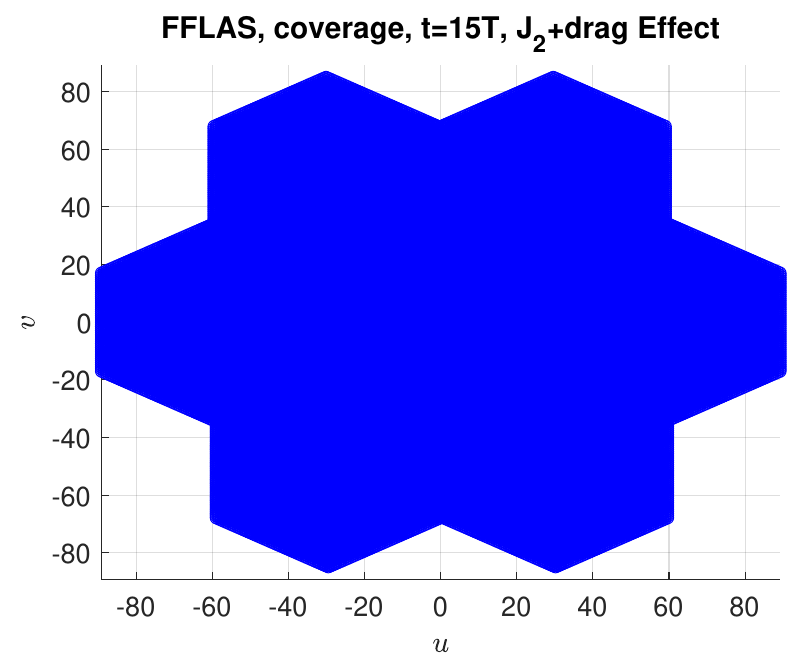}
\label{fig:FFLAS_cov_15T_j2_drag_nocontrol}
}
   \caption{AC and SFC at t=15T for the FFLAS case under $J_{2}$ and drag effect. No control.}
        \label{fig:FFLAS_payload_J2_drag}
\end{figure}

From the simulations in an uncontrolled environment, both orbital and payload considerations are derived. In particular, the satellites oscillate along $x_{N}$ in the unperturbed case. After 0.5 orbital period \textcolor{black}{\textit{T}}, they are mirrored with respect to $x_{N}$ = 0, as shown in Fig. \ref{fig:FFLAS_payload_unperturbed}. If $J_{2}$ effect is present, the satellites are subjected to a growing deviation along $x_{T}$. Oscillation along $x_{N}$ is still there (Fig. \ref{fig:FFLAS_payload_J2}). If, instead, both $J_{2}$ and drag effects are present, satellites are subjected to an $x_{R}$ and $x_{N}$ oscillation (with $x_{R}$ $\textless$ $x_{N}$) and to a drift in $x_{T}$. Moreover, $\dot{x}_{R}$ and $\dot{x}_{T}$ increase in time, so the formation is getting more and more unstable (Fig. \ref{fig:FFLAS_payload_J2_drag}).

From a payload point of view, it has been observed that an in-plane spatial frequency coverage can be obtained from different in-plane mirrored satellite dispositions. This is evident comparing Fig. \ref{fig:FFLAS_coverage} and Fig. \ref{fig:cov_0.5T_nocontrol}. These have, therefore, the same array factor and sidelobe level output (Fig. \ref{fig:FFLAS_af}). For this reason, the unperturbed case shows an average sidelobe level deviation from 0$\%$ up to 18$\%$ with a period of 0.5\textcolor{black}{\textit{T}} (Fig. \ref{fig:FFLAS_payload_unperturbed}). Indeed, a deviation in the spatial frequency coverage or the array factor is visible only if a relative error between the satellites is present, as shown in Fig. \ref{fig:FFLAS_cov_30T_j2_nocontrol}. Therefore, if the entire formation is subjected to a rigid translation, the modelling would still give the same output. This is evident from Fig. \ref{fig:FFLAS_payload_J2_drag}: the significant drift in the $x_{T}$ direction is not visible from the coverage point of view.
If a satellite exhibits a more significant deviation than the others, the spatial frequency coverage and array factor outputs will refer mainly to this error. This is evident in the perturbed cases, in which an $\epsilon_{-2\textcolor{black}{T}}$ behaviour is fairly visible because the second satellite exhibits a more evident deviation along $-x_{N}$ direction (Fig. \ref{fig:FFLAS_payload_J2}). In the perturbed environment, the mean sidelobe level deviation tends to align to 20$\%$ (Fig. \ref{fig:avSLL_j2_period}).

\begin{figure}
\centering
    \subfloat[]{
      \includegraphics[width=0.3\textwidth]{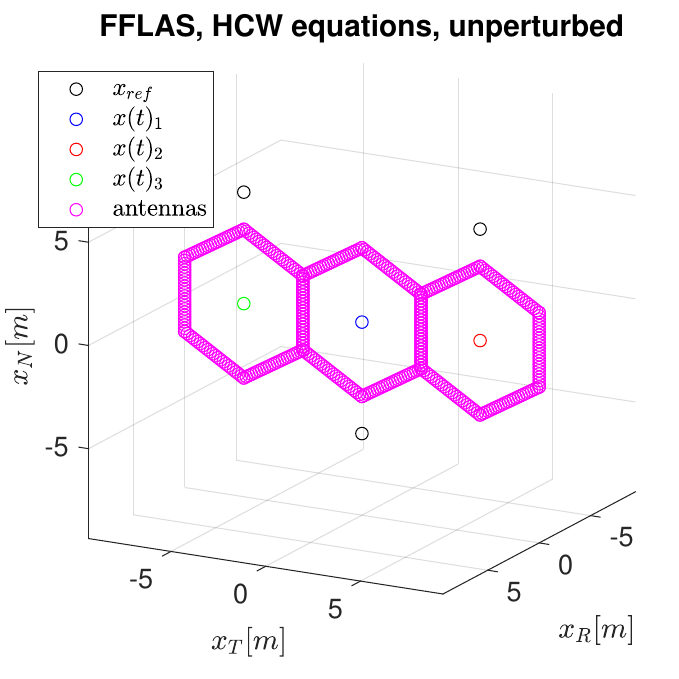}
    }
    \subfloat[]{
      \includegraphics[width=0.3\textwidth]{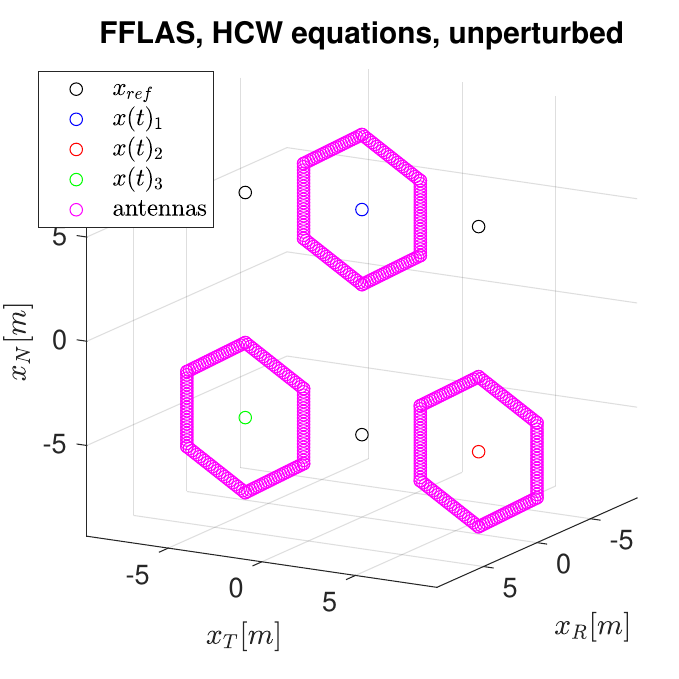}
    }
    \subfloat[]{
      \includegraphics[width=0.3\textwidth]{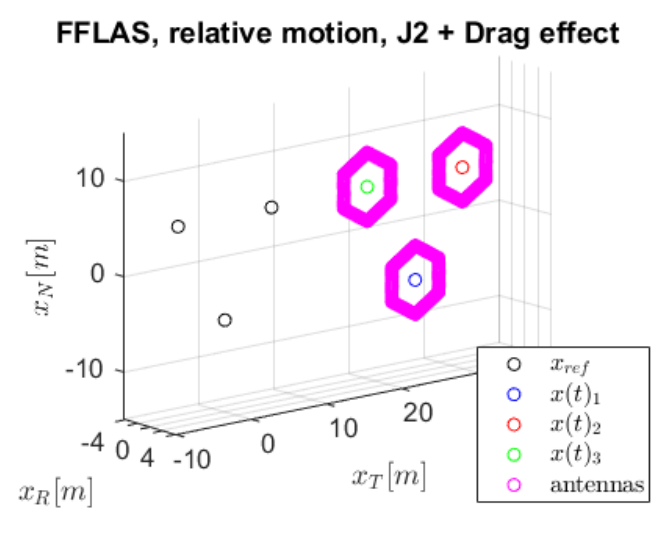}
    }
    \hspace{0mm}
    \subfloat[]{
      \includegraphics[width=0.3\textwidth]{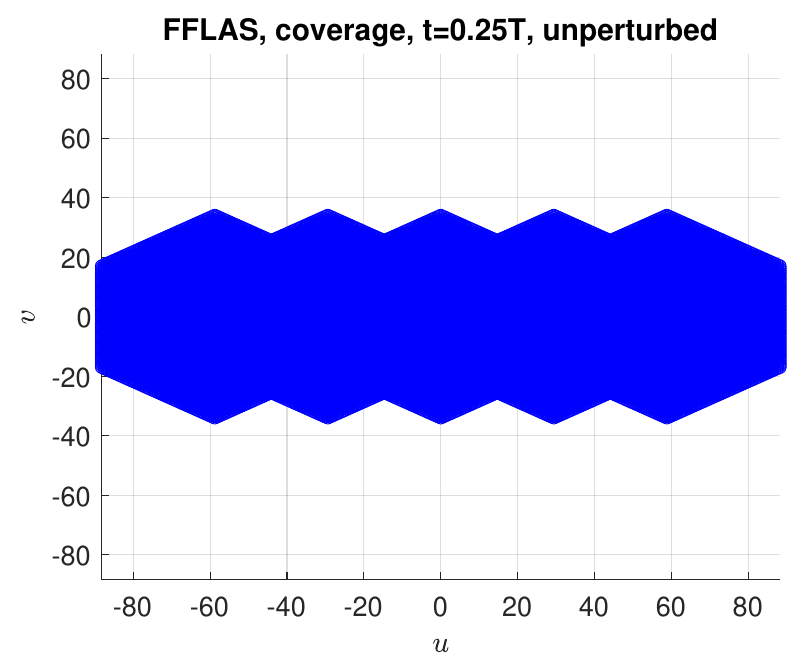}
    }
    \subfloat[]{   
      \includegraphics[width=0.3\textwidth]{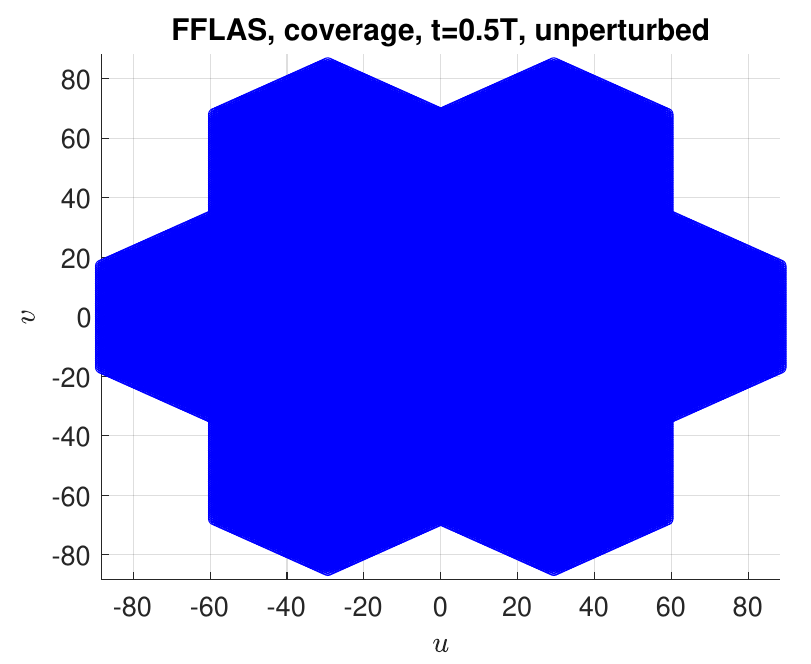}
    }
    \subfloat[]{
      \includegraphics[width=0.3\textwidth]{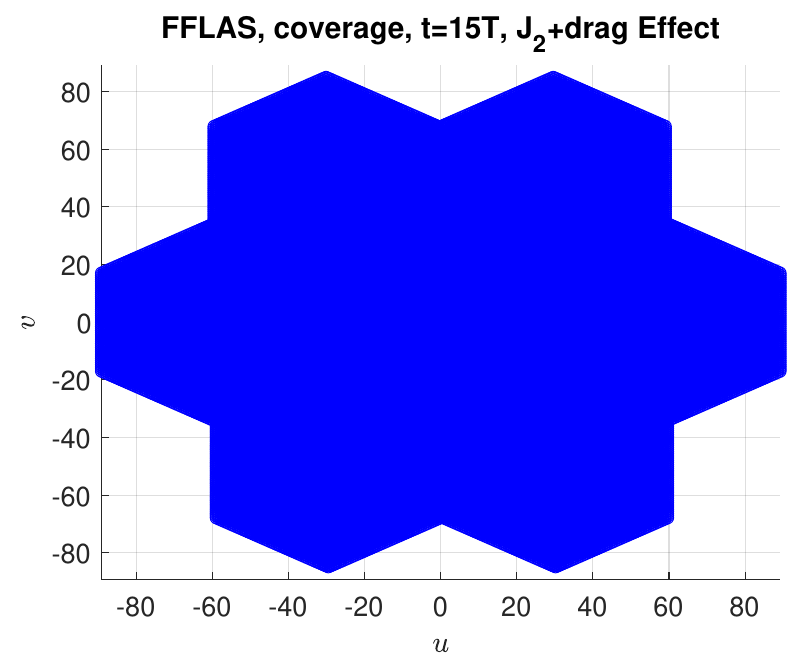}
    }
    \hspace{0mm}
    \subfloat[]{
      \includegraphics[width=0.3\textwidth]{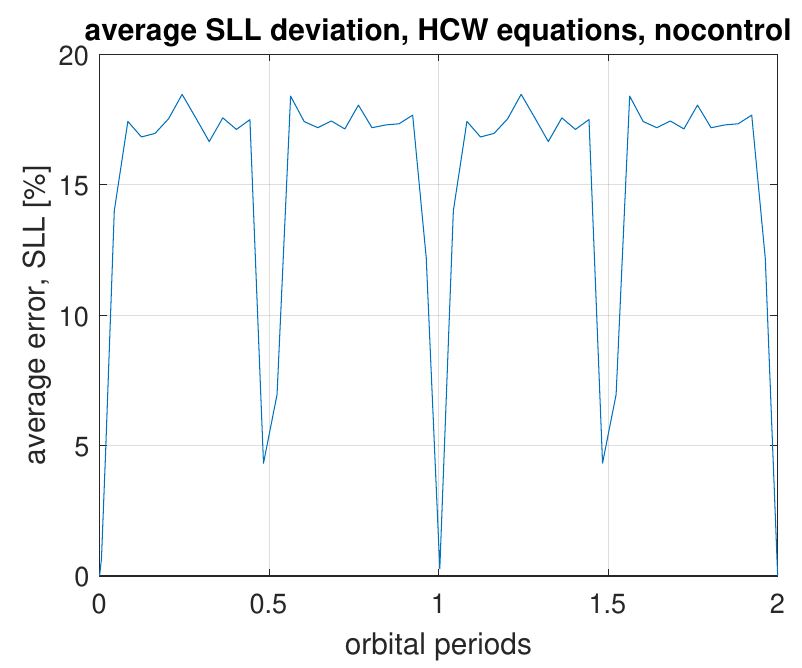}
    }
    \caption{caption}
\end{figure}

\textcolor{black}{As described in \cite{Sabatini2008}, the smaller $\alpha$ the heavier the drift in the transverse direction. Having $\alpha$ = $\beta$ means having the chief satellite equal to the deputy ones: this condition would lead to an only-$J_{2}$ like condition, with a negligible effect of the differential drag. In this environment, $x_{R}$ would show a negligible oscillation compared to the null-$\alpha$ case, and the drift along $x_{T}$ would be way smaller than the analysed case.}
\textcolor{black}{Moreover, considering a virtual chief satellite with null mass and controlling the formation's relative position allows both to act on the relative formation control and on the formation altitude. In this way, the analysis shows how a relative control could keep the right formation altitude and avoid the decay of the absolute orbit.}

\subsubsection{Controlled case} \label{subs:FFLASc}
\begin{figure}[H]
\centering
\subfloat[]
{
      \includegraphics[scale = 0.3]{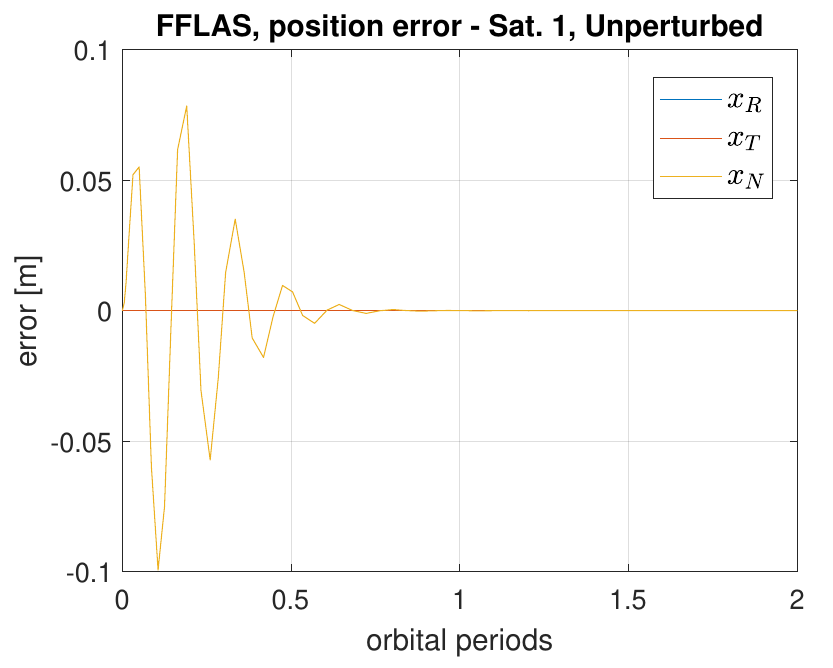}
     }
     \subfloat[]
{
  \includegraphics[scale = 0.3]{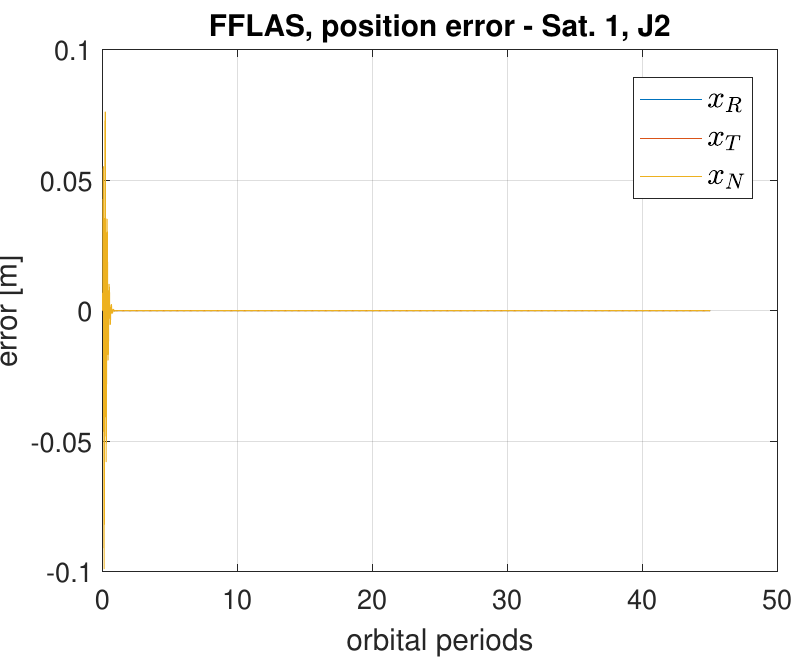}
     }
     
 \subfloat[]{    
     \includegraphics[scale = 0.3]{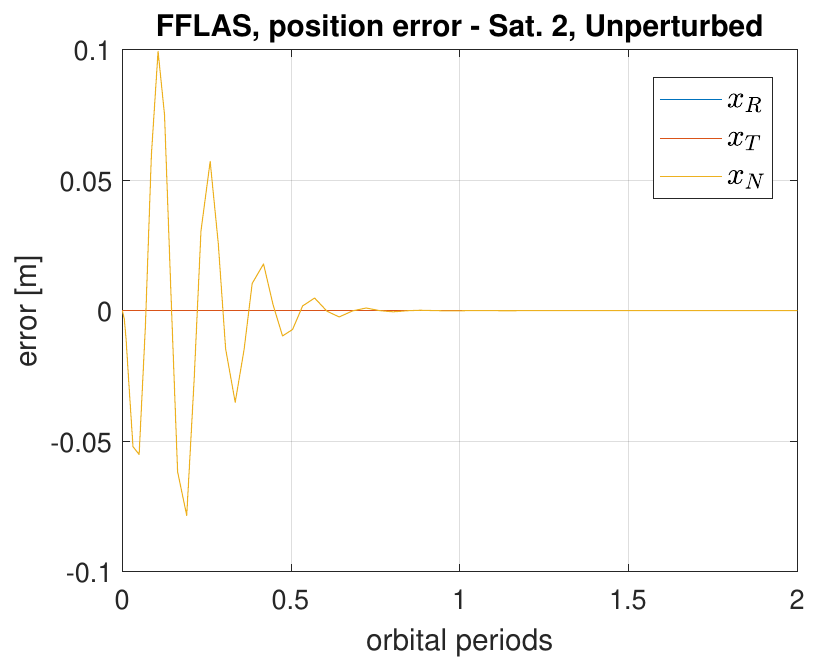}
}
\subfloat[]
{   
 \includegraphics[scale = 0.3]{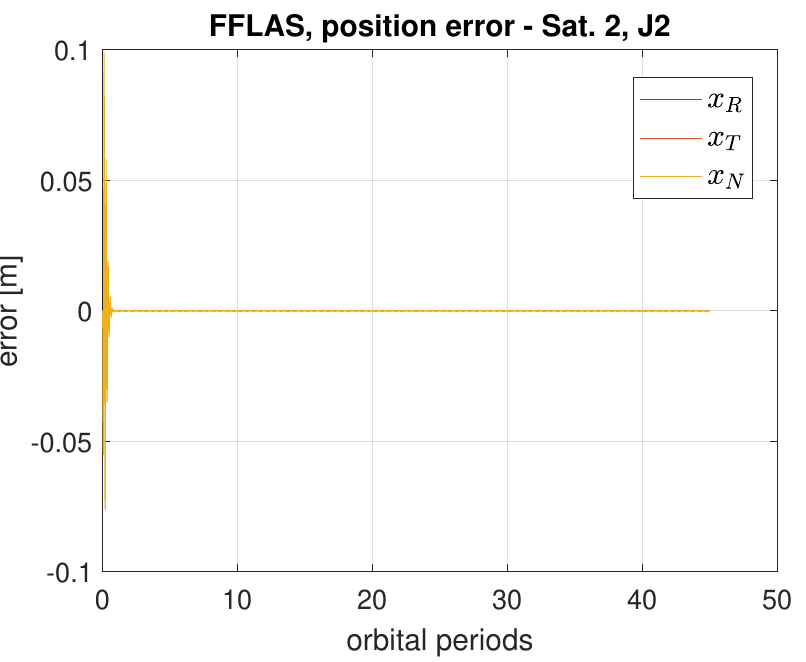}
     }

\subfloat[]{
 \includegraphics[scale = 0.3]{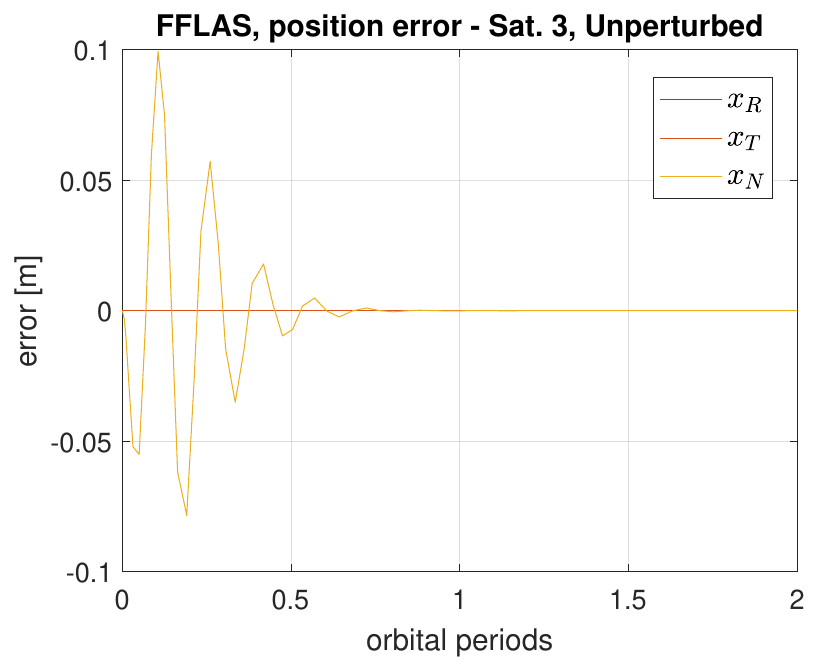}
}
\subfloat[]{
 \includegraphics[scale = 0.3]{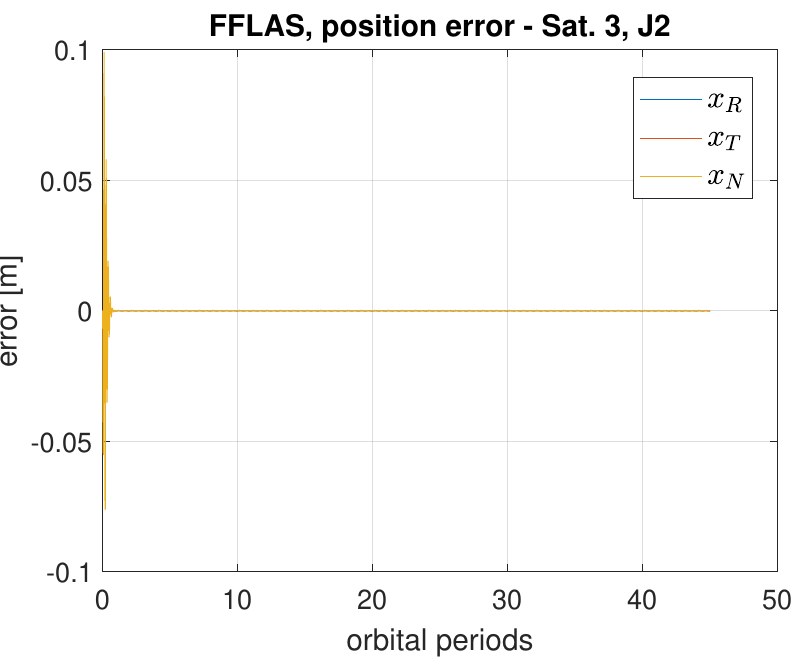}
} 
   \caption{Deviation position error from the nominal state for the FFLAS case both in the unperturbed (left) and under $J_{2}$ effect case (right), PID action.}
        \label{fig:fflas_pos_unpert_j2_control}
        
\end{figure}

\begin{figure}[H]
\centering
\subfloat[]
{
      \includegraphics[scale = 0.3]{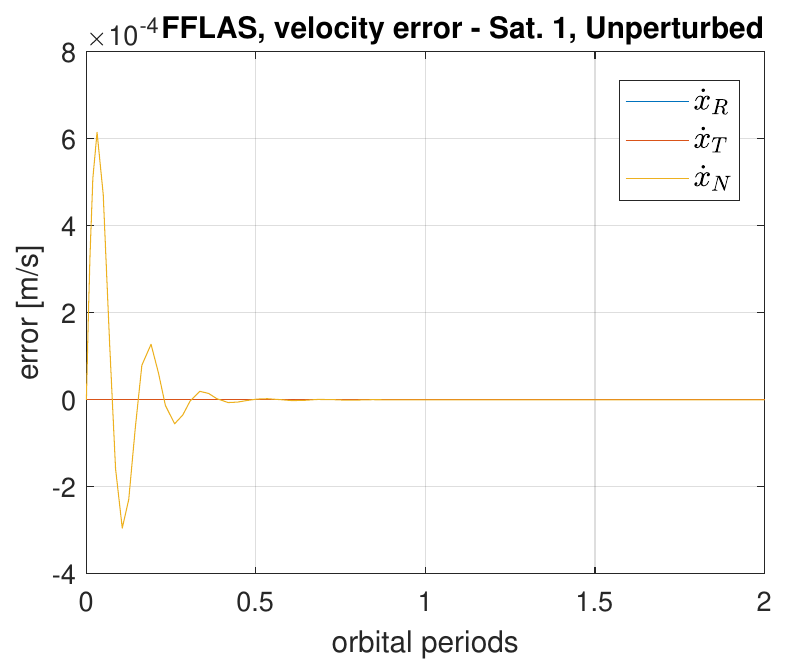}
     }
     \subfloat[]
{
  \includegraphics[scale = 0.3]{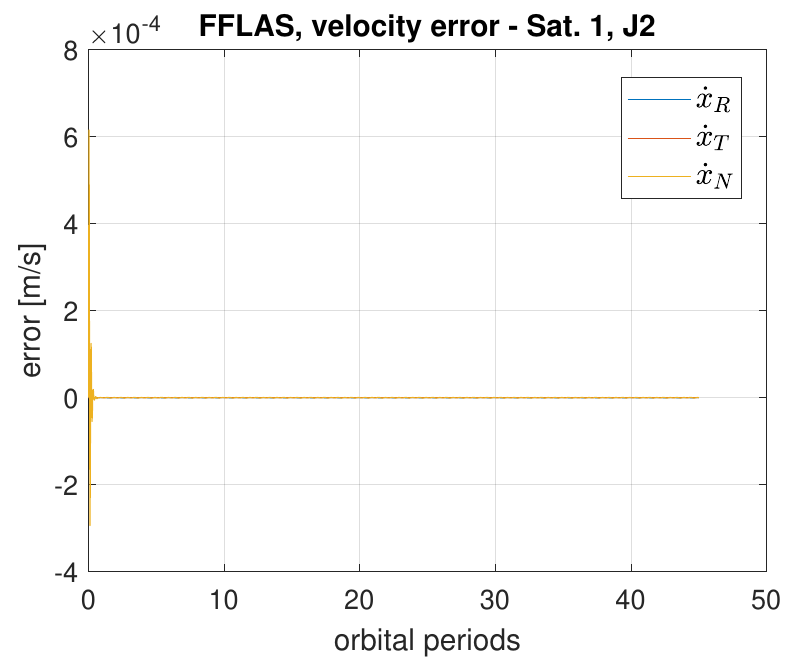}
     }
     
 \subfloat[]{    
     \includegraphics[scale = 0.3]{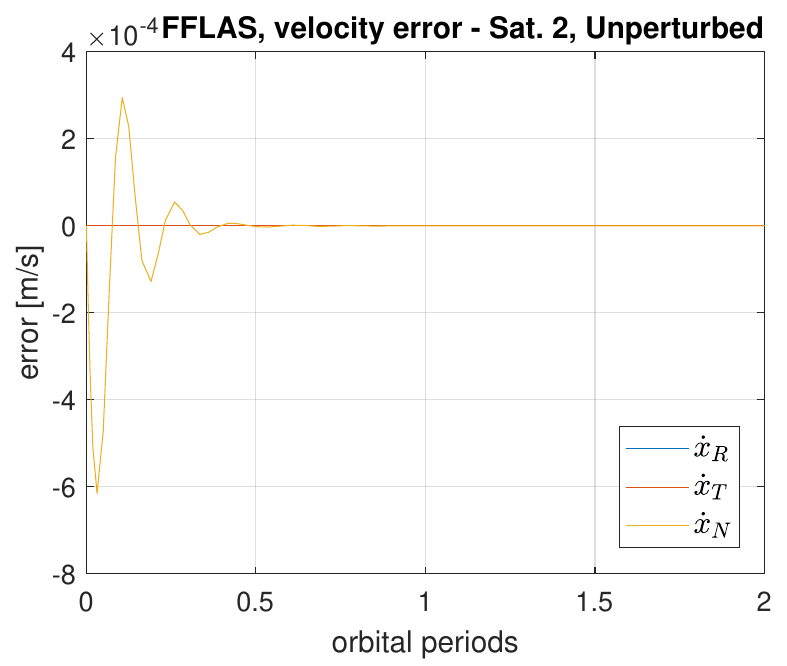}
}
\subfloat[]
{   
 \includegraphics[scale = 0.3]{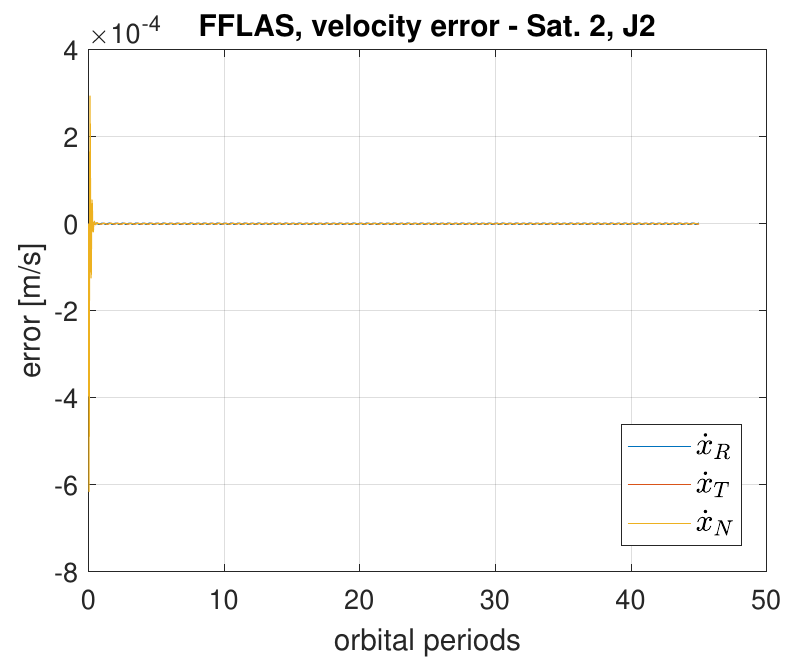}
     }

\subfloat[]{
 \includegraphics[scale = 0.3]{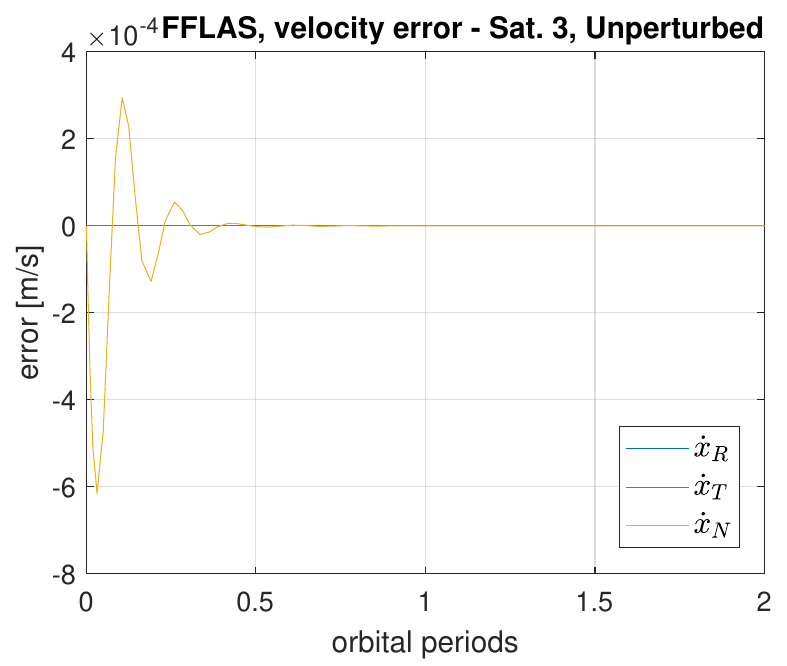}
}
\subfloat[]{
 \includegraphics[scale = 0.3]{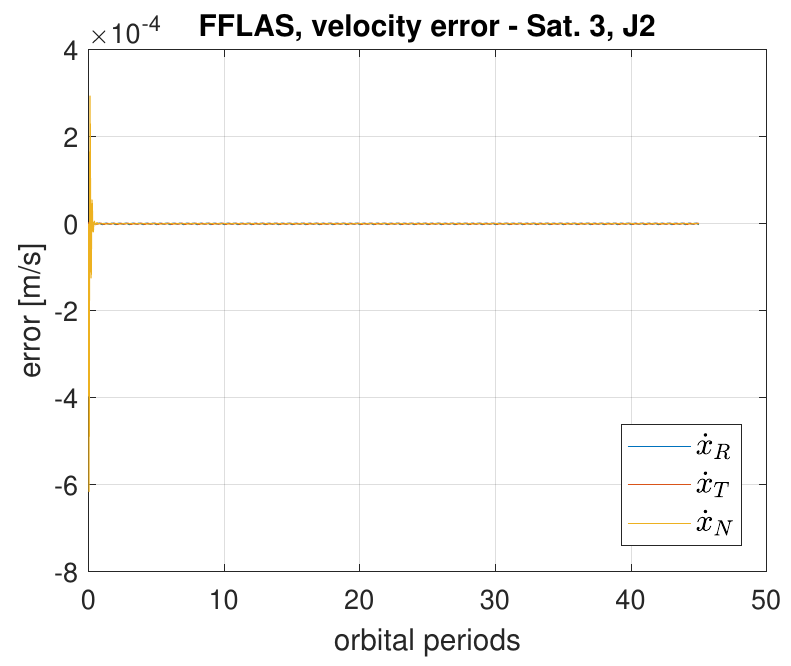}
} 
   \caption{Deviation velocity error from the nominal state for the FFLAS case both in the unperturbed (left) and under $J_{2}$ effect case (right), PID action.}
        \label{fig:fflas_vel_unpert_j2_control}
        
\end{figure}

\begin{figure}[H]
\centering
\subfloat[]{
\includegraphics[scale = 0.3]{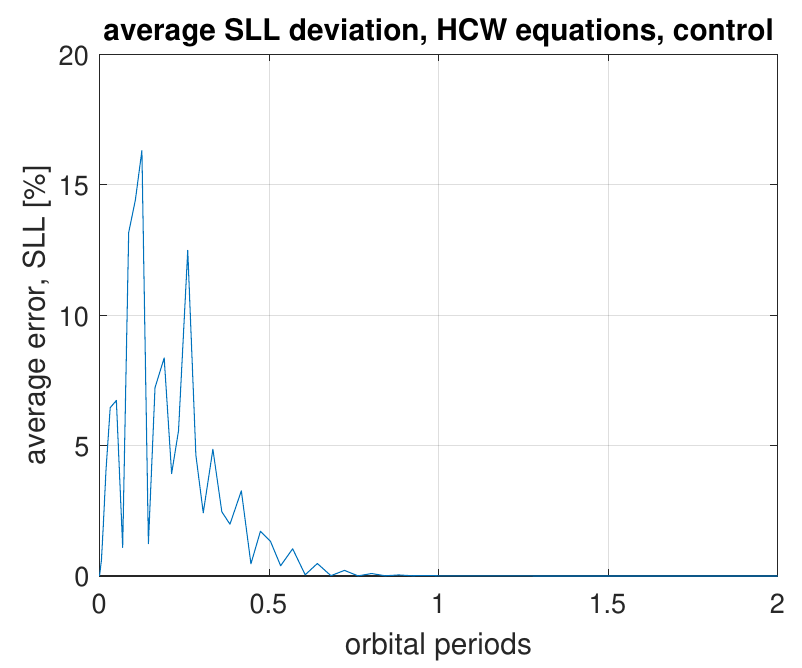}
}
\subfloat[]{
\includegraphics[scale = 0.3]{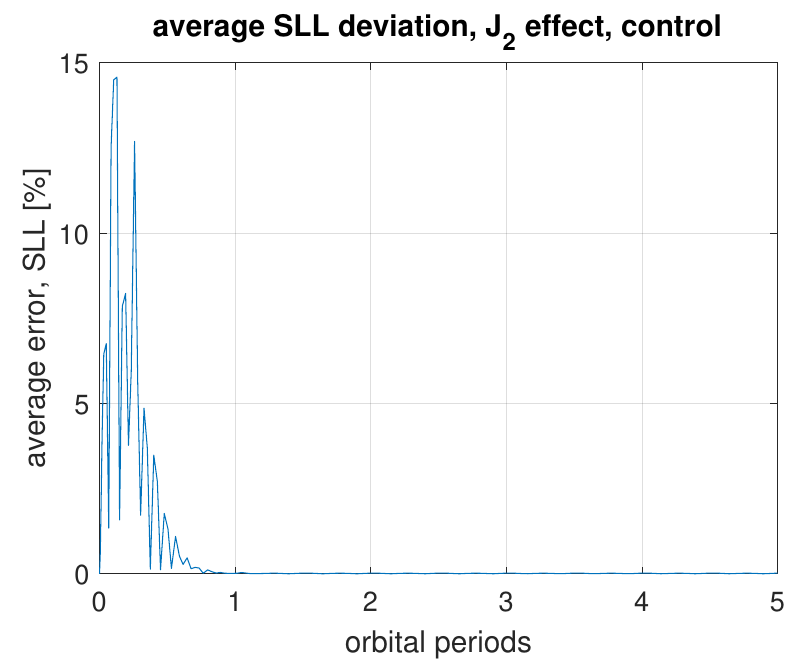}
}
   \caption{Average SLL loss for the FFLAS case  both in the unperturbed (left) and under $J_{2}$ effect case (right), PID action.}
        \label{fig:avSLL_unpert_j2_control}
\end{figure}

A PID controller is implemented to have a control error on the relative position vector below 10 $cm$. This is a linear controller: it is simpler to set and more intuitive, but less effective than a non-linear one. The three controller constants are selected through a trial and error approach. As shown in Table \ref{tab:PIDconstants_influence}, each constant influences differently the instrument impulse response. A proportional response alone does not counteract effectively: a too-small $K_{p}$ ($\sim$ $10^{-6}$) does not stop the $x_{N}$ oscillation, while a higher constant ($\sim$ $10^{-3}$) does but with a final Sts error $\neq$ 0. To modify the Sts error, $K_{i}$ must be introduced. A strict $K_{i}$ ($\sim$ $10^{-3}$) allows to relax the requirements on $K_{p}$ (for instance setting $K_{p}$ = $10^{-6}$); however, a continuous oscillation is observed, so the control is not considered optimal. This is an intrinsic $K_{i}$ characteristic: the integration takes into account the accumulated errors from the past, and can therefore cause the present one to overshoot. An introduction of a $K_{d}$ term might help mitigate this problem.
$K_{i}$ helps also reduce the rise time: it helps therefore setting an error $\leq$ 10 $cm$ from the beginning, the desired condition in a station-keeping simulation. The selected PID constants are depicted in Table \ref{tab:FFLAS_PID}. The simulations in a controlled environment show that the overshoot is $\leq$ 10 $cm$ and the settling time $\leq$ \textcolor{black}{\textit{T}} both for both the unperturbed and the perturbed environment, and the velocity error is around zero in less than one orbit (Fig. \ref{fig:fflas_pos_unpert_j2_control}, \ref{fig:fflas_vel_unpert_j2_control}).
\textcolor{black}{Table \ref{tab:mean_std_J2} shows the mean and standard deviation computation of the position and velocity errors vector e\textit{(t)} for the three satellites in presence of $J_{2}$ only. The computations consider the iterations after one orbit \textit{T}. The results confirm the convergence. Nonetheless, the simulation is performed focusing on an ideal case for the actuators and the navigation errors, which impact considerably on the result.}
The mean sidelobe level deviation goes from 18-20$\%$ in an uncontrolled environment to less than 0.02$\%$ in one orbit (Fig. \ref{fig:avSLL_unpert_j2_control}). The control and the performance of the payload are therefore strictly correlated: correcting the relative deviation implies maximising the performance. 

\begin{table}[H]
\centering
\caption{PID constants.}
\begin{tabular}{ | p{1.5 cm} |p{ 1.5 cm}| p{1.5 cm} |} 
\hline
   \rowcolor[HTML]{C0C0C0} 
 \textbf{$K_{p}$ $ [\frac{N}{m}]$} & \textbf{$K_{d}$ $ [\frac{Ns}{m}]$} & \textbf{$K_{i}$ $ [\frac{N}{ms}]$} \\ \hline \hline
3 $\cdot$ 10$^{-3}$ & 1 $\cdot$ 10$^{-6}$ & 5 $\cdot$ 10$^{-5}$  \\ \hline
\end{tabular}
\label{tab:FFLAS_PID}
\end{table} 

\begin{table}[H]
\centering
\caption{\textcolor{black}{Mean and Standard Deviation values of the position and velocity error after convergence for the three satellites under $J_{2}$ effect. PID action.}}
\begin{tabular}{ | p{0.6 cm}|  p{2.1 cm}|  p{2.1 cm}|  p{2.1 cm}|  p{2.1 cm}|  p{2.1 cm}|  p{2.1cm}|} 
\hline
 \rowcolor[HTML]{C0C0C0}
 \textbf{Sat.} & \textbf{Mean($x_{R}$) [$m$]} & \textbf{Std($x_{R}$) [$m$]}  & \textbf{Mean($x_{T}$) [$m$]} & \textbf{Std($x_{T}$) [$m$]}  & \textbf{Mean($x_{N}$) [$m$]} & \textbf{Std($x_{N}$) [$m$]} \\ \hline \hline
Sat.1 & $6.0455e^{-9}$ &  $3.6232e^{-6}$ &  $-3.1622e^{-9}$ &  $1.5852e^{-6}$ & $6.1876e^{-8}$ & $2.1855e^{-5}$ \\ \hline
Sat.2 &  $9.6244e^{-8}$ & $4.9890e^{-5}$ & $2.6119e^{-8}$ & $3.7562e^{-5}$ & $-6.1928e^{-8}$ & $2.1854e^{-5}$ \\ \hline
Sat.3 & $-1.0834e^{-7}$	& $5.0121e^{-5}$	& $-1.9795e^{-8}	$ & $3.7635e^{-5}$	& $-6.1825e^{-8}$	& $2.1856e^{-5}$ \\ \hline \hline
\rowcolor[HTML]{C0C0C0}
\textbf{Sat.}   & \textbf{Mean($\Dot{x}_{R}$) [$\frac{m}{s}$]} & \textbf{Std($\Dot{x}_{R}$) [$\frac{m}{s}$]}    & \textbf{Mean($\Dot{x}_{T}$) [$\frac{m}{s}$]} & \textbf{Std($\Dot{x}_{T}$) [$\frac{m}{s}$]}    & \textbf{Mean($\Dot{x}_{N}$) [$\frac{m}{s}$]} & \textbf{Std($\Dot{x}_{N}$) [$\frac{m}{s}$]} \\ \hline \hline
Sat.1 & $-3.0181e^{-10}$	& $1.6969e^{-7}$ 	& $-1.3652e^{-10}$	& $7.4156e^{-8}	$& $7.0280e^{-10}$	& $4.9858e^{-7}$ \\ \hline
Sat.2 & $9.8459e^{-10}$	& $1.07566e^{-6}	$ & $-1.6370e^{-9}$	& $8.4627e^{-7}	$ & $ -7.0278e^{-10}	$& $4.9856e^{-7}$  \\ \hline
Sat.3 & $-3.8097e^{-10}$	& $1.1044e^{-6}	$ & $1.9100e^{-9}$	& $8.5401e^{-7}$	& $-7.0282e^{-10}	$& $4.9861e^{-7}$ \\ \hline
\end{tabular}
\label{tab:mean_std_J2}
\end{table}

\textcolor{black}{In this work we have considered an ideal control without taking into account noises and uncertainties typical of the on-board thrusters and actuators. Also, the control scheme is demanding from a fuel point of view, as it involves a continous control against the natural oscillation of the relative motion in the normal directions, as also described in \citep{Scala2023}. A different approach to exploit natural solution of the relative motion to reduce the fuel consumption for formation maintenance has been studied in \citep{Neira_2023_TriHex}. However, this concept has not been included in this work, which aims at correlating the payload output with formation position errors. }

\section{Conclusion} \label{sec:conclusion}
The work presented in this article wants to analyse the modelling and control of a formation flying with an L-band payload, taking the FFLAS study as a reference configuration. The study shows that there exists an intrinsic bond between a control error (affecting position and velocity) and the payload output. A deeper examination could help when performing a configuration trade-off or refining the control action necessary for formation keeping.

The study confirms the potential of a hexagonal array compared to other closed or open ones, and opens the path to radio frequency interference reduction, a critical aspect for current interferometry missions. Moreover, it investigates the array factor computation in two different ways, both for results validation and redundancy optimization. However, the focus is made on the ideal case. 

The position error sensitivity analysis performed on FFLAS gives a preliminary draft of how the payload changes due to a position error, and it detects some kind of symmetry between the input error and the resultant payload output. In particular, an approach between satellites is more critical from a sidelobe level point of view than an estrangement. The dynamics of a satellite and its payload can be combined, and interesting results can be obtained. For instance, this primary sensitivity analysis opens the path to improving orbit prediction accuracy. A machine learning approach could be developed to have an alternative way to physics-based models and improve the orbit prediction accuracy. In particular, the machine learning approach could be taught to learn from several array factor outputs, so to associate each one to the initial configuration the satellites are into, and recognise where the error is coming from. This article has not tested this idea: its feasibility should be verified in a future analysis.

This work opens the path to a formation flying error sensitivity analysis, with the credit of focusing not only on a control perspective but especially on a modelling one. However, many simplifications are applied, such as considering only the highest sidelobe, or few data. 
A more probabilistic approach is followed, introducing some innovative performance parameters like the \textit{P} matrix. This approach improves the single-error, single-direction one, as it introduces an automated process and considers all in-plane errors. However, many limitations are applied. Few runs are evaluated, and they are set in a conservative approach. Moreover, only in-plane errors are considered.
In both single and random cases, the analyses could be completed with out-of-plane errors or with rotation ones and extended to multiple errors in all directions, \textcolor{black}{following a refined model of the actual dynamics of the formation. In this sense,} the examination shall include the velocity. 

Finally, the study delineates the theory behind the dynamics and the control of a formation flying in low Earth orbit, introducing the $J_{2}$ and drag perturbations through a linearised model and a PID controller for the station-keeping action.
This approach could be further investigated. 
The method could analyse other attitude components, such as the sensors and the actuators, and add other evaluations. For instance, the analysis could focus on the required $\Delta v$ from the thrusters or perform a plume impingement analysis, \textcolor{black}{as the plume impingement could generate severe degradation effects on optical payloads and spacecraft materials, as explained in  \cite{Scala2023}}.  
\textcolor{black}{The control action shall consider its fuel consumption request, taking into account the thrusters' capacity, as explained in \cite{Scala2023}.}

The natural dynamic is analysed also from a modelling point of view. This is an innovative perspective as, usually, control focuses only on the position and velocity outputs. Some modelling parameters are displayed to verify how the payload reacts to the formation flying dynamics. Some of the considerations from the position error sensitivity analysis are here found. As a main result, the study shows that the major impact on the modelling output is due to a relative error between the satellites. 
In future work, the sidelobe level could be linked to other payload parameters, for instance, finding a maximum allowable sidelobe level deviation for imaging requirements. Once this value is found, the PID constants could be further refined so to respect this maximum value.

\section{Acknowledgments}
This article is part of the COMPASS project: 'Control for orbit manoeuvring by surfing through orbit perturbations' (Grant agreement No 679086). This project is European Research Council (ERC) funded project under the European Union's Horizon 2020 research (www.compass.polimi.it).

\bibliographystyle{model5-names}
\biboptions{authoryear}
\bibliography{refs}


\end{document}